\font\gital=cmti10 scaled \magstep4
\font\logo=logo10 scaled \magstep2

\documentstyle[12pt,twoside]{article}

\begin{document}

%
%

\begin{titlepage}     
\vphantom{x}\hfill Revised manuscript \\
hep-ph/9510305 \hfill November 1996 
   \vspace{2cm}  \begin{center} {\large \bf 
Thermal variational principle and gauge fields
   } \\  \vfil  \vspace{2cm} {\large 
York Schr\"oder }  \footnote{ 
e--mail : yschroed@x4u2.desy.de}  \\[6pt]
DESY, Theory Group, Notkestr. 85, D--22603 Hamburg, Germany \\
   \vfil \vspace{1cm} {\large 
Hermann Schulz }   \footnote{ 
e--mail : hschulz@itp.uni--hannover.de} \\[6pt]
Institut f\"ur Theoretische Physik, Universit\"at Hannover,
Appelstr. 2, D--30167 Hannover, Germany \\
   \end{center}  \vfil \vspace{2.2cm}
%
%
   \begin{quotation} \ \ 
A Feynman--Jensen version of the thermal variational principle is
applied to hot gauge fields, Abelian as well as non--Abelian$\,$:
scalar electrodynamics (without scalar self--coupling) and the
gluon plasma. The perturbatively known self--energies are shown
to derive by variation from a free quadratic (''Gaussian'') trial
Lagrangian. Independence of the covariant gauge fixing parameter
is reached (within the order $g^3$ studied) after a reformulation
of the partition function such that it depends on only even
powers of the gauge field. Also static properties (Debye
screening) are reproduced this way. But because of the present
need to expand the variational functional, the method falls short
of its potential nonperturbative power.

    \vspace{.3cm} \noindent
PACS number(s) : \ 11.10.Wx \ and \ 11.15.Tk 
    \end{quotation}

\vfil \vspace{.4cm}
\end{titlepage}

%
%
   
\vspace{1.5cm} \centerline{\bf 
I. \ INTRODUCTION } \vspace{1cm} 

Variational principles, well established in nonrelativistic
quantum problems, develop their true power by setting a measure
for the best approximate solution within some parametrization
of a trial space. This space is made up of wave functions
$\psi$ in non--relativistic quantum mechanics, by statistical
operators $\rho$ in thermodynamics, and by actions $S$ in the
Feynman--Jensen formulation. The above object $\psi$ can always
be understood to be the ground state of some Hamiltonian $H$.
Any statistical operator is also related uniquely to a Hermitean
operator $H$ (its ''Hamiltonian''). Thus, in any of these three
cases, we may talk about a {\sl theory} to be parametrized and
varied. The task is to find a class of theories coming reasonable
close to the truth but to keep it simple enough for tractability.

Each of the above three cases generalizes to field theory, $\psi$
becoming a wave functional, see e.g.~\cite{wavef}, while $H$ and
$S$ keep their meaning. For the formulations of the thermal
variational principle, with $H$ [2--10] or with $S$ [5,10--15],
we refer to the next Section as some part II of the Introduction.

We focus on the application to pure gauge theory with particular
interest in the hot gluon system. There are three fascinating
aspects of this system. First, it distills out from the standard
model by reducing the number of flavours to zero, while possibly
still containing the whole non--Abelian mystery. Second, other
than in the Higgs sector (the other distillate) its Lagrangian
looks so simple$\,$: ${\cal L} = - F^2 / 4\,$. Third, its
high--temperature limit may be called understood, mainly based on
the rearrangement \cite{BP,effa} of diagrams at soft--scale outer
momenta $Q$ (key words$\,$: resummation, hard thermal loops).

There are several basic problems and pitfalls at the very
beginning when the variational method contacts gauge fields.
It is the subject of the present paper to make the calculus
working {\sl at all}. Hence no new results on the hot gluon
system should be expected here. Instead, the known perturbative
results are used to {\sl test} our new variational construction.
With this first step we hope to pave the road towards its
presumedly powerful nonperturbative possibilities.

Specifically on SU($N$) gauge fields under thermal variation,
there is (to our knowledge) only the one (thus pioneering) paper
of R. Manka \cite{mank} in 1986. He studied pure non--Abelian
gauge theory by using a free trial Lagrangian, namely the Abelian
one for photons taken $n$--fold ($n \equiv N^2-1$). The fields are
identified, $A_{\rm non-Abelian}=A_{\rm trial}$ (at least in the
high--temperature phase), and a constant transverse photon mass
is taken as variational parameter. Note that this identification
of pure, but interacting gauge fields with free, but massive ones
makes the trial theory nontrivial. With the longitudinal mass
included as well as the 4--vertex (which both were neglected in
\cite{mank}), Manka conjectures that the perturbative results
on masses, generated by the plasma, should be obtained from
variation as well. Indeed, they should -- but by this supposedly
easy task we were led into all that follows.

The following outlook reflects, to some extent, our individual
path into the subject. Starting with the basic ideas of
\cite{mank} just mentioned, we were more or less forced into one
step (away from \cite{mank}) after the other$\,$:
\begin{enumerate}
\item The covariant gauge--fixing parameter $\alpha$ is
      reintroduced, and kept arbitrary, because all experience
      with, e.g., the damping puzzle of the gluon plasma tells us
      that $\alpha$, if surviving in final results, is an ideal
      indicator for wrong physics.
\item Both dynamical mass terms (transverse and longitudinal)
      are included as functions of momentum. This setup covers
      static screening as well as dynamically generated masses.
      The massive--photonic trial theory still keeps its Abelian
      gauge invariance.
\item The functional integral formulation is applied. In passing,
      although our notation is Minkowskian (metrics $+ - - -$)
      we actually always {\sl mean} the Euclidean space. We only
      have to remember, at appropriate places, that the zeroth
      component $A^0$ of the gauge field is $i$ times a real field.
\item The classical (or Feynman--Jensen) version of the variational
      functional is used, because it avoids difficulties in
      constructing the Hamiltonian to our higher--derivative
      trial--Lagrangian. As an intermediate result, the covariant
      gauge--fixing parameters of studied and trial theory become
      equal.
\item The variational functional, if evaluated with the quadratic
      photonic trial theory as described, still depends on the
      (common) gauge--fixing parameter $\alpha$ (as also observed
      in \cite{ibst}). A way out is proposed by first rewriting
      the partition function of the theory studied such that its
      action becomes even in the gauge fields. This is called
      the ''even version'' in Sec.~II C.
\item As in the low--order perturbative treatments, and since we
      shall only reproduce its results, detailed renormalization
      is not (yet) required in this paper. Divergent terms can
      be separated from the finite thermal ones. Hence, the
      coupling $g$ changes its meaning to be the running
      coupling in these thermal contributions.
\item For a first application of the ''even version'', scalar
      electrodynamics is appreciated once more \cite{sed} to be an
      ideal toy model for the non--Abelian problem. The known
      self--energies are put in by hand, but supplied with variable
      prefactors. Through variation, the latter become 1 indeed.
\item In the non--Abelian case, the Faddeev--Popov determinant
      becomes part of the even--odd decomposition. The ''even''
      functional works well, except for a (hopefully) minor
      detail at the end (concerning gauge--fixing dependence in
      higher order).
\item For the explicit analysis just mentioned, the variational
      functional had to be expanded up to the third (partly
      fourth) $g$--power. This apparently inevitable recourse
      to $g$--powers is a big disappointment.
\end{enumerate}

The paper is organized as follows. Section II on the formulations
of the thermal variational functional is a continued Introduction.
Especially the ''even version'' (the one that works) is
introduced in Sec.~II C. In Section III we follow the
Feynman--Jensen version. It leads to unphysical results, but is,
on the other hand, reasonable simple to introduce several
technical details. Section IV treats scalar electrodynamics with
the ''even version'' of the functional. In Section V on the gluon
plasma, things start more involved but become very similar at the
end. In Section VI the case of constant trial self--energies is
discussed in terms of Debye screening and magnetic mass. Open
questions are summarized in Section VII. Conclusions follow in
Section VIII. Three Appendices cover details on the functional
integral measures, on some normal integrals involved and on sum
rules.

%
%

\vspace{1.5cm} \centerline{\bf
II. \ THE THERMAL VARIATIONAL PRINCIPLE } 

\vspace{.7cm} \centerline{\bf 
II A. \ Gibbs --- Bogoljubov } \vspace{.4cm}

The extremal properties of thermodynamic potentials are known
from text books \cite{huan,bren} on statistical physics. In
particular in the canonical ensemble (the only one considered
in this paper), the free energy takes its minimum at
equilibrium$\,$: $F\ge F_\bullet$. In its usual version, the
thermal variational principle is identical with this modest
inequality, if its left--hand side is detailed$\,$:
\begin{equation} \label{2v}
  \hbox{{\gital v}\,} \,\left[\, H \,\right]\, \;\;\equiv
  \;\;\, {\rm Tr}\, \left( {e^{-\beta H} \over  Z}
  \,\left[\, H_\bullet - H \,\right]\, \right) - T
  \ln \left( Z \right) \;\;\; {\buildrel ! \over = } \;\;\;
  \mbox{min.} \;\; . \;\;
\end{equation}
The proof is given shortly. In (\ref{2v}) and in the following
an index bullet refers to the system studied (at equilibrium), 
i.e. to the ''hard problem'' which one likes to learn about by
the variational method. $\beta = 1/T$, $Z=\, {\rm Tr}\,
\left( e^{-\beta H} \right)$. Trial quantities carry no index, so
$H$ is the element running through the trial space whose only
restrictions are that (a) the spectrum of $H$ is bounded from
below and (b) $H$ acts in the Hilbert space of $H_\bullet$. The
formulation (\ref{2v}) is found e.g. as eq.~(10.83) in
\cite{huan} or as eq.~(20.37) in \cite{tjab}. It is called Gibbs
variational priciple in \cite{huan} and Bogoljubov inequality in
\cite{mank,tjab}. There is a natural application to the
Heisenberg spin model, where mimimizing {\gital v}\, yields
the best Curie--Weiss Hamiltonian \cite{tjab}, thereby justifying
the mean field procedure.

For the proof of (\ref{2v}), we claim that one line suffices. It
rests on the inequality $-\ln (x)\ge 1-x$ and on the irrelevance
of operator--ordering under trace, $\, {\rm Tr}\, \ln (AB) = \,
{\rm Tr}\, \ln (BA)$. With any non--equilibrium statistical
operator $\rho$, the line reads$\,$:
\begin{equation} \label{2ineq}
  F \left[ \rho \right] \; = \;\, {\rm Tr}\, \left( \rho
  H_\bullet \right) + T \, {\rm Tr}\, \left( \rho \ln\,
  \left[\, \rho \,\right]\, \right) \;\; = \;\; F_\bullet - T
  \, {\rm Tr}\, \left( \rho \ln \,\left[\, {1\over \rho} \,
  \rho_\bullet \,\right]\, \right) \;\;\ge \;\; F_\bullet
  \;\; . \;\;
\end{equation}
To the left, (\ref{2ineq}) starts with the non--equlibrium free
energy in ($E-TS$)--form. The knowledge of $\rho_\bullet$ at
equilibrium, $H_\bullet = - T \ln \left(\rho_\bullet Z_\bullet
\right)$, is used for the inner equality sign ($F_\bullet
= - T \ln \left( Z_\bullet \right)$). Finally, the right end has
been simplified using $\, {\rm Tr}\,\left(\rho\right) =
\,{\rm Tr}\, \left( \rho_\bullet \right) = 1$.
In a whatsoever non--equilibrium state the system is, it has a
statistical operator $\rho$ with the three properties 1--trace,
hermitecity and positivity. Thus (with the properties (a), (b)
of $H$ as stated above), its general form is $\rho = e^{-\beta H}
/ Z$. This makes (\ref{2ineq}) to become (\ref{2v}), ~q.~e.~d.

It is tempting to require that the trial theory be a solvable one
(e.g. a free field theory). However, it must not. Imagine there
was a small coupling $e$ in the trial theory, and (for simplicity)
only one variational parameter $\eta$. Near its minimum, the
functional would take the form {\gital v}$\; = a(e) + b(e)
\,\left[\, \eta - c(e) \right]^2$. Clearly, through perturbative
expansion of {\gital v}\,, the coefficients $a$, $b$, $c$ as
well as the position $\eta$ of the minimum would be obtained as
power series in $e$. The parameter $\eta$ may be chosen to be
the coupling $e$ itself.

We now turn to gauge field theory, governing a periodically
repeated box of volume $V$ and coupled to a thermal bath at
rest with four--velocity $U=(1,{\bf 0} )$. In the variational
principle (\ref{2v}) $H_\bullet$ and $H$ are the Hamiltonians to
a Lagrangian ${\cal L} _\bullet$ studied and a trial Lagrangian
${\cal L}$, respectively. To count the same number of field
degrees of freedom, one may either prepare the physical Hilbert
spaces from the outset \cite{mank,almu,york} or work with
extended spaces (corrected by ghosts). In the latter case the
two gauge--fixings may be different. Adopting general covariant
gauges, there is a gauge fixing parameter $\alpha_\bullet$ of
the theory studied and an $\alpha$ of the trial theory. No final
result is allowed to depend on either of them.

The hot gluon system is described by the pure Yang--Mills
Lagrangian
\begin{equation} \label{2lbu}
 {\cal L}_\bullet \; = \; - {1 \over  4} \,
 {F_\bullet}_{\mu\nu}^{\,\enskip a} F_\bullet^{\mu\nu \; a}
 - {1 \over  2 \alpha_\bullet} \,\left( \partial^\mu \!
 A_\mu^a \right)^2 + \, i \overline{c}^{\, a} \partial^\mu
 D_\mu^{ab} c^b \quad . \;\;
\end{equation}
with ${F_\bullet}_{\mu\nu}^{\,\enskip a} = \partial_\mu A_\nu^a
- \partial_\nu A_\mu^a + g f^{abc} A_\mu^b A_\nu^c$ and
$D_\mu^{ab} = \delta^{ab} \partial_\mu - g f^{abc} A_\mu^c\,$.
In the high--temperature limit the 1--loop contributions (hard
thermal loops) are of relative order unity and therefore must be
included in the true leading order \cite{BP,effa}. The gluon
propagator, resummed this way, may be written as
\begin{equation} \label{2g}
  G^{\mu\nu} (Q) = {\hbox{\logo A}^{\mu\nu} (Q) \over
  Q^2 - M_t(Q) }  + {{\sf B}^{\mu\nu} (Q) \over
  Q^2 - M_\ell(Q) } + \alpha \; {{\sf D}^{\mu\nu} (Q) \over
  Q^2} \;\; , \;\;
\end{equation}
where $\hbox{\logo A}$, ${\sf B}$, ${\sf D}$ are members of
the symmetric Lorentz matrix basis
\begin{eqnarray} \label{2ad}
 & & \hspace{-1cm} \hbox{\logo A} = g - {\sf B} - {\sf D} \;\; ,
 \;\;{\sf B}= { V\circ V\over  V^2 }\;\; ,\;\; {\sf C} =
  { Q \circ V + V \circ Q \over
      \sqrt{2\,}^{\hbox to0.2pt{\hss$
      \vrule height 2pt width 0.6pt depth 0pt $}\;\!}
  Q^2 q } \;\; , \;\; {\sf D}= { Q \circ Q \over  Q^2 } \\
  \label{2v4}
  & & \hspace{-1cm}  \mbox{with} \quad V= Q^2 U-(UQ)Q=(-q^2\, ,
  \, -Q_0{\bf q} \, ) \;\; . \;\;
\end{eqnarray}
The orthonormal properties of (\ref{2ad}) are listed in
\cite{sumr}. Note that $\hbox{\logo A}$ and ${\sf B}$ are
projectors$\,$:
\begin{equation} \label{2pro}
  \hbox{\logo A}^{\mu\nu}(Q)\; Q_\nu = 0\quad , \quad
  {\sf B}^{\mu\nu} (Q)\; Q_\nu=0 \;\; . \;\;
\end{equation}
In (\ref{2g}), $M_t = {\mit\Pi}_t$ and $M_\ell = {\mit\Pi}_\ell$
are the well known polarization functions \cite{kapu,lawe}
\begin{equation} \label{2pi}
  {\mit\Pi}_t(Q) = {3\over 2} m^2 - {1\over 2}{\mit\Pi}_\ell(Q)
  \quad , \quad {\mit\Pi}_\ell (Q) = 4 g^2 N \sum_P \Delta_0^{-}
  \Delta_0 \,\left[\, p^2 - { \left({\bf p} {\bf q} \right)^2
  \over  q^2 } \,\right]\, \;\;
\end{equation}
with
\begin{equation} \label{2mds}
  m^2 = {g^2 N T^2 \over  9} \;\; , \;\;\Delta_0 = {1\over
  P^2} \;\; , \;\; \Delta_0^{-} = { 1 \over  (Q-P)^2 }\;\;  ,
  \;\; \sum_P \equiv {1\over V} \sum_{\bf p} T \sum_n \;\;
\end{equation}
For more details on the ${\mit\Pi}$'s (especially in our
notation) see Appendix B of \cite{nt}. If $V\to\infty\,$,
$\sum_P$ turns into $\int\! d^3p (2\pi)^{-3} T \sum_n\,$. We
work with the Matsubara contour$\,$: $Q = (i\omega_n, {\bf q}
)\,$, $\omega_n=2\pi nT$. The gauge fields are Fourier
transformed as
\begin{equation} \label{2fou}
  A_\mu (x) = \sum_P e^{-iPx} A_\mu (P) \quad , \quad
  A_\mu (P) = \int^\beta e^{iPx} A_\mu (x) \quad \;\;
\end{equation}
with $x=(-i\tau , {\bf r} )$ and $\int^\beta \equiv \int_0^\beta
\! d\tau \int\!d^3r$. To e.g. check this, the thermal Kronecker
symbol
\begin{equation} \label{2kron}
  \int^\beta\! e^{i (Q-P)x} = \beta V \delta_{n_Q, n_P}
  \,\delta_{{\bf q} , {\bf p}} \;\; \equiv \;\; \,\left[\,
  Q - P \,\right]\,  \quad
\end{equation}
is very convenient. In (\ref{2g}) the bullet on $\alpha_\bullet$
had been ''forgotten'', because the Greens function (\ref{2g})
will turn out to be that of the trial theory as well. 

Were there not the paper \cite{mank}, a suitable trial Lagrangian
could come into mind while contemplating on (\ref{2g}). Use $n$
free photon Lagrangians (numbered by $a$), supply them with
variable mass terms such that (\ref{2g}) is among their
propagators, and identify the fields$\,$:
     $A$ in (\ref{2lbu}) $\equiv A$ in (\ref{2l})$\,$:
\begin{equation} \label{2l}
  {\cal L} = - {1\over 4} \, F^2 + {1\over 2} A \left( MA \right)
  - {1\over 2\alpha} \left( \partial  A\right)^2 \;
  + \; i \overline{c} \partial^2 c \;\;
\end{equation}
with $F_{\mu\nu} = \partial_\mu A_\nu - \partial_\nu A_\mu$.
The trivial index $a$ is suppressed here, and
\begin{equation} \label{2m}
   \left( MA \right)^\mu (x) = \int^\beta_{x^\prime} \,\sum_Q
   e^{-iQ(x-x^\prime )} \,\left[\, M_t(Q)
   \hbox{\logo A}^{\mu\nu} (Q) + M_\ell(Q) {\sf B}^{\mu\nu} (Q)
   \,\right]\, \, A_\nu (x^\prime ) \;\; . \;\;
\end{equation}
The propagator of (\ref{2l}) is (\ref{2g}). But note that we are
still free to choose e.g. constant masses
\begin{equation} \label{2coma}
  M_t = m_t^2 \quad , \quad  M_\ell = m_\ell^2
  \qquad \mbox{(''$m$--case'')} \;\; , \;\;
\end{equation}
or to cover the true leading--order propagators (\ref{2g}) with
\begin{equation} \label{2lamb}
  M_t = \lambda_t^2 {\mit\Pi}_t(Q) \quad , \quad M_\ell
      = \lambda_\ell^2 {\mit\Pi}_\ell(Q)
  \qquad \mbox{(''$\lambda$--case'')} \;\; . \;\;
\end{equation}

Our trial Lagrangian (\ref{2l}) is non--interacting and quadratic
in the fields $A$. The gauge--fixing term is necessary, because
the mass terms are Abelian gauge invariant. To see this, insert
the Fourier transform (\ref{2fou}) into (\ref{2m}) and notice
that the gauge variation $\delta A_\nu(Q) = - Q_\nu\,\chi (Q) $
drops out due to (\ref{2pro}). Despite these neat properties of
the mass term, the longitudinal one makes trouble. By $Q^\mu\to
i\partial^\mu$ in (\ref{2m}), and in the $m$--case for simplicity,
we may rewrite $A(MA)$ as
\begin{equation} \label{2ama}
  A (MA) = A_\mu (x) \,\left[\, m_t^2\, \hbox{\logo A}^{\mu\nu}
  (i\partial ) + m_\ell^2\, {\sf B}^{\mu\nu} (i\partial )
  \,\right]\, A_\nu (x) \;\; , \;\;
\end{equation}
In passing, (\ref{2ama}) is the Abelian (trivial) case of the
Lagrangian considered in \cite{kreu}. The matrix ${\sf B}$ (not
$\hbox{\logo A}$) has a denominator containing $Q_0\,$: $V^2
= -q^2 Q^2 \to - \Delta \Box$. Hence, our trial Lagrangian has
arbitrarily high powers in the time derivative. The definition
of field momentum densities in higher derivative Lagrangians is
a delicate matter, as is the construction of its Hamiltonian.
Thus $H$, the trial object, makes the problem. Note that, if
working with a constant--mass Stueckelberg term ${1\over 2} m^2
A^\mu A_\mu$ \cite{itzu}, this problem would not yet arise.
We leave these difficulties right now, because there is a
wonderful way out as detailed in the following subsection.

\vspace{1cm} \centerline{\bf 
II B. \ Feynman --- Jensen } \vspace{.4cm}

To each Lagrangian, (\ref{2lbu}) and (\ref{2l}), there is a
partition function which, using functional integrals, is
expressed by the actions $S_\bullet = - \int^\beta
{\cal L} _\bullet$ and $S= - \int^\beta {\cal L}\;$:
\begin{equation} \label{2z}
   Z_\bullet\; =\; {1\over Z^\bullet_B}\; {\cal N} \int\!
   {\cal D} A \; e^{-S_\bullet} \quad , \quad Z\; = \; {1
   \over Z_B}\; {\cal N} \int\! {\cal D} A \; e^{-S} \;\; . \;\;
\end{equation}
In (\ref{2z}), and for the moment, let $\int {\cal D} A $
include the ghost field integrations. (\ref{2z}) holds true in
Euclidean space \cite{bern,kapu}$\,$: $A_0$ is a purely imaginary
field. The prefactors $Z^\bullet_B$ and $Z_B$, e.g.
$Z_B = \int {\cal D} B \, e^{-\int^\beta B^2/2\alpha}$, occur
through the derivation of (\ref{2z}) while integrating over
$\delta (\partial A -B)$ with normalized weight. Usually,
they are hidden
in the functional measure ${\cal N}$. But here, the two
${\cal N}$ in (\ref{2z}) are equal and independent of $\alpha$.
Nevertheless, they depend on $\beta$ \cite{bern}. It might be
emphasized, that the above functional language can still be
applied to the Hamiltonian version {\gital v}\,$\left[\,
H\,\right]\,$, since the first term of (\ref{2v}) is
$ -T \ln \left( Z \right)$, and the individual terms in
$\left\langle H_\bullet -H\right\rangle$ ({\,\sl if known\,})
can be related to Greens functions, which in turn derive from
$Z$ (with source terms included).

Hamiltonians can be avoided at all, as we learn in \S ~3.4 of
Feynman's text book \cite{feytext} (see also \S ~8.3,4 there
and \cite{fey54}). Start from $Z_\bullet$, add the factors
$e^{S}$ and $e^{-S}$ under the integral, divide by (and multiply
with) $Z$ and define the average $\left\langle\ldots\right\rangle$
as given in (\ref{2zge}). Then,
\begin{equation} \label{2zge}
   Z_\bullet \;\; =\;\; {Z_B\over  Z^\bullet_B}\; Z\; \left\langle
   e^{- \left( S_\bullet - S \right)} \right\rangle \;\;\ge \;\;
  {Z_B\over  Z^\bullet_B}\; Z\; e^{- \left\langle S_\bullet
  - S \right\rangle } \quad , \quad \left\langle \ldots
  \right\rangle \equiv { \int {\cal D} A e^{-S} \ldots
  \over \int {\cal D} A e^{-S} } \;\; . \;\;
\end{equation}
The above inequality is, in the case at hand, the Jensen
inequality, see e.g. \cite{blat}. Its simplest version states
that $\left\langle e^{-f} \right\rangle\ge e^{- \left\langle f
\right\rangle}$. The proof rests on the convexity of $e^x$
\cite{chai,klei}, or, equivalently, on the nice figure 3.5 of
\cite{feytext}. For convenience we take the logarithm of
(\ref{2zge})$\,$,
\begin{equation} \label{2vv}
  \hbox{{\gital v}\,} \,\left[\, S \,\right]\,
  =  F + T \left\langle S_\bullet - S \right\rangle
     - T \ln \left( {Z_B \over Z^\bullet_B} \right) \;\;\;
  {\buildrel ! \over = } \;\;\; \mbox{min.} \quad , \;\;
\end{equation}
and call (\ref{2vv}) the Feynman--Jensen variational principle.
The last term, we come back to shortly, is obviously specific
to gauge theory.

In the non--Abelian case, there is a terrible pitfall hidden in
(\ref{2zge}). Admittedly, things were written down, to run into
it with ease. If one still reads $\int {\cal D} A$ to include
the ghost field integrations (wrong case), the ghost term would
appear in the average $T\left\langle S_\bullet - S\right\rangle$
and loose the term linear in $A$, hence all $A$--dependence. But
if the integration over ghosts is correctly recognized to be the
Faddeev--Popov determinant, FP($A$), it may be included into
$S_\bullet$ as $S_\bullet^{\,\rm no\; ghosts} -\ln ({\rm FP}(A))$.
Now, in the avergage $T\left\langle - \ln ({\rm FP})
\right\rangle$, even powers of $A$ survive. For the explicit
formulation of this see Sec.~V. We learn that $i\overline{c} c$,
though being Hermitean, must not be viewed as a real number.
Hence, in the non--Abelian case one needs to write the action
$S_\bullet$ in (\ref{2zge}) and (\ref{2vv}) as to include
FP($A$), while there will be no ghost field integrations in
$\int {\cal D} A$.

There is also the Peierls' version \cite{peie,huan,tjab,chai} of
a thermal variational principle. It states that $\, {\rm Tr}\,
e^{- \beta H} \ge \sum_n e^{- <n\vert H\vert n>}$, rests on
Jensen and may be used for another derivation \cite{chai} of
(\ref{2v}). Things are closely related. But we have no rigorous
answer to the question, whether the two versions (\ref{2v}) and
(\ref{2vv}) are identical statements -- just formulated in
different language -- or not (see also pt.3 in Sec.~VII). The
Feynman--Jensen variational principle stays useful even at
zero temperature. It has been applied at $T=0$ to $\lambda\phi^4$
theory without \cite{ibpo} and with gauge sector \cite{ibst}.
Formerly, these efforts had a Hamiltonian formulation, considered
even thermal \cite{hast}. At finite temperature, but without the
exotic last term, (\ref{2vv}) has been recently used to obtain
gap equations in lattice $\phi^4$--theory \cite{kerr}. Before,
it played a central role in a study of spin models and lattice
gauge theory \cite{rueh}.

The role played by the unusual last term in (\ref{2vv}) clears
up by combining it with the gauge--fixing terms contained in
$S_\bullet$ and $S\;$:
\begin{eqnarray} \label{2gau}
   \hbox{{\gital v}\,}_{\rm gauge} &=&
    {T\over 2} \left( {1\over \alpha_\bullet}
   - {1\over \alpha}\right) \int^\beta \left\langle
   (\partial A)^2 \right\rangle \; - T \ln \left(
   {\prod_{{\bf q} , n}}^\prime
\sqrt{{\alpha \over \alpha_\bullet}\,}^{\hbox to0.2pt{\hss$
\vrule height 2pt width 0.6pt depth 0pt $}\;\!}
    \right)  \nonumber \\
  &=& \left( {\alpha \over \alpha_\bullet} - 1 - \ln \,\left[\,
  {\alpha\over \alpha_\bullet} \,\right]\, \right)
  {V\over 2}\, {\sum_Q}^\prime  \;\; . \;\;
\end{eqnarray}
For the logarithm in the first line see (\ref{azb}), the prime
excludes $n={\bf q} =0$. To understand the last term in the
second line, remember that $\sum_Q = (T/V) \sum_n \sum_{\bf q}$.
For the other terms insert (\ref{2fou}) and use
\begin{equation} \label{2aa}
  \left\langle A_\mu (Q)\, A_\nu (P)\right\rangle = \,\left[\, Q
  + P \,\right]\,\; G_{\mu\nu} (Q)
  = \,\left[\, Q + P\,\right]\,\left( \hbox{\logo A}\Delta_t +
  {\sf B}\Delta_\ell +\alpha{\sf D}\Delta_0 \right) \;\; , \;\;
\end{equation}
where the shorthand notation should be obvious from (\ref{2g}).
Now consider $\alpha$ of the trial Lagrangian to be one of the
variational parameters. Clearly, with respect to $\alpha$,
(\ref{2gau}) has an extremum at $\alpha = \alpha_\bullet$, and
{\gital v}\,$_{\rm gauge}$ vanishes at this position.
Moreover, it is a minimum, since the blank sum at the end in
(\ref{2gau}) is positive (though quartic divergent). By far the
best $\alpha$ is $\alpha_\bullet$. We note three consequences
of $\alpha=\alpha_\bullet$. First, the three terms selected in
(\ref{2gau}) may be simultaneously omitted in the sequel.
Second, there is still dependence on the now common $\alpha$, as
it enters through $\left\langle\ldots\right\rangle$ when traced
back to the trial propagator (\ref{2g}). Hence, the above
selection of $\alpha$--dependent terms was incomplete. But the
divergence of the last factor in (\ref{2gau}) helps maintaining
the conclusion with rigour. Third, with respect to $\alpha$, the
variational principle is exhausted, so one should no more
think about an ''optimal'' (common) $\alpha\,$.

\vspace{1cm} \centerline{\bf 
II C. \ The \ ''even version''} \vspace{.4cm}

So far, we were able to circumvent the Hamiltonian dilemma noted
at the end of Sec.~II A. But in the new version (\ref{2vv})
there is again a troubling element, as we become aware of next.
Terms odd in the gauge field $A$ (the $AAA$ part of
${\cal L}_\bullet$ in particluar) drop out in {\gital v}\,
entirely, because they only enter $\left\langle S_\bullet
- S \right\rangle$ and vanish there, since the average weight is
the quadratic trial action. It is as if the 3--vertex was taken
out from the outset. But a Yang--Mills theory with no 3--vertex
can never be tested suitably by any trial--theory. For more
details see the next Section.

For the resolution to this puzzle, it appears that the usual
philosophy (''improve the trial theory'') fails. Also, our trial
theory (\ref{2l}) is physically so reasonable$\,$: it ''must''
work. Our way out is to introduce one more version of the
variational functional. On one hand this construction, which we
call the ''even version'', is the decisive success in treating
gauge fields variationally. On the other hand the idea is rather
simple$\,$: in general, odd--in--$A$ terms in the action can be
avoided from the outset by playing around with the functional
integrations over $A$ as follows.

Let us split the action into $S_\bullet = {\cal E} + {\cal O}$
with ${\cal E}$ keeping and ${\cal O}$ changing sign under
$A \to -A$. The same decomposition can be done with the
exponentiated action as $e^{- {\cal E}} e^{- {\cal O}}
= e^{- {\cal E}} {\rm ch} ({\cal O}) - e^{- {\cal E}} {\rm sh}
({\cal O} )$. Since the second term drops out under the
functional integrations over the gauge field $A$, we may write
\begin{equation} \label{2dec}
  \int {\cal D} A \; e^{-S_\bullet} \; = \;  \int {\cal D} A \;
  e^{- {\cal E} \; + \; \ln \,\left[\, {\rm ch}
  \left( {\cal O} \right) \,\right]\, } \;\; . \;\;
\end{equation}
The new exponent, which we call $ - S_{\bullet\bullet}\,$, is an
even functional of $A$. Since the above steps precede the use
of Jensen's inequality, quite a new functional {\gital v}\,
arises$\,$:
\begin{equation} \label{2vvv}
  \hbox{{\gital v}\,} \,\left[\, S \,\right]\, = F
  + T \left\langle S_{\bullet\bullet} - S \right\rangle \;\;\;
  {\buildrel ! \over = } \;\;\; \mbox{min.} \;\quad \mbox{with}
  \quad S_{\bullet\bullet} = {\cal E} - \ln \,\left[\, {\rm ch}
  \left( {\cal O} \right) \,\right]\, \;\; . \;\;
\end{equation}
In (\ref{2vvv}) $\alpha=\alpha_\bullet$ is understood, i.e. the
logarithm of $Z_B$'s is omitted together with the gauge fixing
terms in ${\cal E} $ and $S$.

Once there are only even terms in the theory studied, the
quadratic trial theory has a good chance to reproduce the
leading--order perturbative results. We shall show in Sections IV
and V that the ''even version'' works that way, indeed. There,
the Faddeev--Popov determinant (depending on $A$ in Sec.~V, but
not in Sec.~IV) is part of $S_\bullet$ and hence subject of the
above ''even''--ing procedure.

%
%
   
\vspace{1.5cm} \centerline{\bf
III. \ TRIAL AND ERROR } \vspace{1cm}  

In this short Section we step back to the insufficient
Feynman--Jensen formulation (\ref{2vv}) to see which way it goes
wrong, to introduce some basic integrals and for a first run
through the necessary algebra in the simplest case.
For simplicity, let us even omit the Faddeev--Popov term (i.e.
run into the pitfall noticed below (\ref{2vv})). It is not
(solely) responsible for the defect, as we shall remark at
the end of this Section.

Using $\alpha=\alpha_\bullet$ as reasoned below (\ref{2aa}), the
functional reads {\gital v}$\; = F + T \left\langle
S_\bullet - S \right\rangle$. In the difference $S_\bullet - S
= -\int^\beta \left( {\cal L} _\bullet - {\cal L} \right)$
the terms odd in the gauge field $A$ vanish under the average
$\left\langle\ldots\right\rangle$. Others cancel. The only two
surviving terms are
\begin{equation} \label{3s-s}
   T \left\langle S_\bullet - S \right\rangle = V {1\over 2}
   \left\langle A^a(MA^a) \right\rangle + V {g^2\over 4}
   f^{abc}f^{ars} \left\langle A_\mu^b A_\nu^c  A^{\mu\, r}
   A^{\nu\, s} \right\rangle \;\;\equiv \hbox{{\gital v}\,}_M
   \; + \; \hbox{{\gital v}\,}_{AAAA} \;\; , \;\;
\end{equation}
where $\int^\beta$ has reduced to $\beta V$ due to
spacetime--independence of the averages. The first term,
{\gital v}\,$_M$ with $M$ from (\ref{2m}), is readily
evaluated by using (\ref{2fou}), (\ref{2aa}) and the trace
relation $(\,\left[\, M_t \hbox{\logo A} + M_\ell {\sf B}\,\right]
\, G\,)_\mu^{\;\;\mu} = 2 M_t \Delta_t + M_\ell \Delta_\ell\;$:
\begin{equation} \label{3vm}
  \hbox{{\gital v}\,}_M = nVT^4 \left( - L_t
  - {1\over 2} L_\ell \right)  \;\; . \;\;
\end{equation}
$L_{t,\ell}$ are two sums out of the collection$\,$:
\begin{equation} \label{3sums}
  J_{t,\ell} = -\beta^2 \sum_P \Delta_{t,\ell} \;\; , \;\;
  L_{t,\ell} = -\beta^4 \sum_P P^2 \left(\Delta_{t,\ell}
               - \Delta_0 \right)\;\; ,\;\;
  Y_{t,\ell} = \beta^2 \sum_P p^2 \Delta_0 \Delta_{t,\ell} \;\;
\end{equation}
with $\Delta_{t,\ell} = 1/(P^2-M_{t,\ell}(P))$ and $\Delta_0 =
1/P^2$. The prefactor $n$ in (\ref{3vm}) comes from the trivial
sum over the colour index.

The treatment of {\gital v}\,$_{AAAA}$ starts with
the Wick decomposition \cite{mank} of the average into three
pairs with partners
\begin{equation} \label{3aa}
  \left\langle A_\mu^a (x) A_\nu^b (x) \right\rangle
  = \delta^{ab} \sum_P G_{\mu\nu} (P) \; = \; \delta^{ab}\,
  T^2\, {1\over 3} \left( u_{\mu\nu} r + v_{\mu\nu} s
  \right) \;\; . \;\;
\end{equation}
The first equality in (\ref{3aa}) derives with (\ref{2fou}),
(\ref{2aa}). The second one arises after integration over the
directions of ${\bf p}$. As the propagators $\Delta_{t,\ell}$
are rotationally invariant (even in the $\lambda$--case
(\ref{2lamb})), this angular integration amounts to the
replacements $\hbox{\logo A} \to -{2\over 3} u$, ${\sf B} \to
- {1\over 3}u - {1\over 3} p^2 \Delta_0 v$ and
${\sf D} \to {1\over 3} (v-u) + {1\over 3} p^2 \Delta_0 v$
with the Lorentz matrices $u_{\mu\nu}$, $v_{\mu\nu}$ given by
$U_\mu U_\nu - g_{\mu\nu}$ and $4U_\mu U_\nu - g_{\mu\nu}$,
respectively. For the sums $r$ and $s$ see (\ref{3rs}) below.
Using the first equation (\ref{3aa}) and with $f^{abc}f^{abc}
= nN$ one derives the first line of (\ref{3aaaa}). Exploiting
the $u$--$v$--version, one arrives at the second one$\,$:
\begin{eqnarray} \label{3aaaa}
  \hbox{{\gital v}\,}_{AAAA} &=& n V {g^2 N \over 4}
  \left( \;\left[ \sum_P G_\mu^{\;\;\mu} (P) \right]^2
   - \sum_P G_{\mu\nu} (P) \sum_Q G^{\mu\nu} (Q) \, \right)
  \nonumber \\
  &=& nVT^4\, {g^2 N\over 6} \left( r + s \right)
  \left( r - 2s \right) \;\; . \;\;
\end{eqnarray}
The objects $r$, $s$ in (\ref{3aa}),(\ref{3aaaa}) are given by
\begin{equation} \label{3rs}
  r = 2 J_t + J_\ell + \alpha J_0 \quad , \quad
  s = - Y_\ell - \alpha J_0 + \alpha Y_0 \;\; , \;\;
\end{equation}
where $J_0$ and $Y_0$ are the sums of (\ref{3sums}) taken
at vanishing mass.

The last term of {\gital v}\, to be evaluated is the trial
free energy $F=-T\ln(Z)$. First of all, since colours do not
mix, $Z$ is an $n$--fold product$\,$,
\begin{equation} \label{3f}
   F = n\, F_{\rm colourless}  = - n T \ln \left(
   Z_{\rm colourless} \right) \; = \; n\, V\, T^4\,\left(
   - 2 I_t - I_\ell + I_0 \right) \;\; , \;\;
\end{equation}
and the colourless partition function is identical with that
of scalar ED, see Sec.~IV, if omitting the factor due to the
scalars. In the formula (\ref{2z}) for $Z$ (read colourless and
Euclidean), there are three unknown flying objects$\,$:
${\cal N}$, $\int{\cal D} A$ and $\int{\cal D} B$. This is not
a shame if $Z$ is used exclusively as a generating functional.
But here we need $Z$ as a precise number. The trouble \cite{bern}
with the normalization factor ${\cal N}$ is proportional to the
care of its treatment. We make efforts in Appendix A to write
down at least (if not to derive) this factor ${\cal N}$. Here we
take from (\ref{aend}) that $F$ indeed splits up into the terms
in the right--hand side of (\ref{3f}). With (\ref{an0}), we
obtain
\begin{equation} \label{3fl}
   I_\ell = {1 \over 2 V T^3} \sum_{\bf p} \,\left[\,
   \ln \left( - T^2 \Delta_\ell(P_0=0, {\bf p} ) \right)
   + {\sum_n}^\prime \ln \left( P_0^2
   \Delta_\ell(P) \right) \,\right]\, \;\; , \;\;
\end{equation}
where the prime excludes $n=0$. The index $\ell$ may be replaced
by $t$ or by $0$ (then referring to zero mass). The expression
(\ref{3fl}) sticks with this awkward form as long as the
$\lambda$--case (\ref{2lamb}) is included. But by differentiation
with respect to $\lambda_\ell$ we may write
\begin{equation} \label{3ilrel}
  - \lambda_\ell \partial_{\lambda_\ell} I_\ell = L_\ell
    \qquad \mbox{or} \qquad I_\ell = I_0
    - \int_0^{\lambda_\ell} \! d\lambda \, {1\over \lambda} \,
    L_\ell (\lambda_\ell = \lambda ) \;\; , \;\;
\end{equation}
the right half being equivalent to a coupling constant
integration. In the $m$--case the above relation reads
$- m \partial_m I = L$.

The sums $I$, $J$ to $Y$ are divergent, and one has to keep
track of variational--parameter dependences while renormalizing
\cite{mank}. To study this in simple terms (and for the rest of
this Section) we turn to constant masses by (\ref{2coma}). In
this case the frequency sum in (\ref{3fl}) can be done
\cite{kapu}. Using (\ref{af4}) and going to the infinite volume
limit, one obtains
\begin{equation} \label{3i}
   I_\ell = - {1\over 2 \pi^2} \int_0^\infty\! dx\, x^2
   \,\left[\, {1\over 2}
\sqrt{{x^2 + \varepsilon_\ell^2}\,}^{\hbox to0.2pt{\hss$
\vrule height 2pt width 0.6pt depth 0pt $}\;\!}
   + \ln \left( 1 -  e^{-
\sqrt{{x^2 + \varepsilon_\ell^2}\,}^{\hbox to0.2pt{\hss$
\vrule height 2pt width 0.6pt depth 0pt $}\;\!}
   } \right)  \,\right]\,  \quad , \quad
   \varepsilon_\ell \equiv \beta m_\ell \;\; . \;\;
\end{equation}
Furthermore, $L_\ell = \varepsilon_\ell^2 J_\ell\,$. The sum $J$
becomes
\begin{equation} \label{3j}
   J_\ell = {1\over 2\pi^2} \int_0^\infty\! dx \, {x^2 \over
\sqrt{{x^2 + \varepsilon_\ell^2}\,}^{\hbox to0.2pt{\hss$
\vrule height 2pt width 0.6pt depth 0pt $}\;\!}
   } \,\left[\, {1\over 2} + {1 \over e^{
\sqrt{{x^2 + \varepsilon_\ell^2}\,}^{\hbox to0.2pt{\hss$
\vrule height 2pt width 0.6pt depth 0pt $}\;\!}
   } - 1 } \,\right]\, \;\;
\end{equation}
with clearly the ${1\over 2}$--term being UV--divergent as in
(\ref{3i}). Even after subtracting zero--point energies by hand
(which the functional integral does not know of),
$I_\ell\to I_\ell + {1\over 4\pi^2}\int_0^\infty\! dx\, x^3
\equiv I_\ell^{\rm sub}$, there remains a singular integral
depending on the variational parameter $\varepsilon_\ell$.
On the other hand, in a low--order perturbative treatment,
such terms can be addressed as zero--temperature renormalization
\cite{BP,nt} and omitted entirely. As we like to reproduce
these results only, the omission should be allowed here as
well. Consider, for example, the combination
$ -I_\ell^{\rm sub} - {1\over 2} L_\ell$, which occurs in
$F + \hbox{{\gital v}\,}_M$, and supply $p$ with an
UV--cutoff $\Lambda\;$:
\begin{eqnarray}
   \,\left[\, -I_\ell^{\rm sub} - {1\over 2} L_\ell \;
   \right]_{{1\over 2}{\rm -term}}
   &=& {1\over 4\pi^2} \int_0^{\Lambda /T} \! dx \left(
\sqrt{{x^2+\varepsilon_\ell^2}\,}^{\hbox to0.2pt{\hss$
\vrule height 2pt width 0.6pt depth 0pt $}\;\!}
   - x - {\varepsilon_\ell^2 \over 2
\sqrt{{x^2+\varepsilon_\ell^2}\,}^{\hbox to0.2pt{\hss$
\vrule height 2pt width 0.6pt depth 0pt $}\;\!}
   } \right) \qquad  \nonumber \\  \label{3comb}
   &=& {\varepsilon_\ell^4 \over 32\pi^2} \ln\left(
   {\Lambda\over T\varepsilon_\ell}\right)
   \;\; + \;\; O \left( \varepsilon_\ell^4 \right) \;\; . \;\;
\end{eqnarray}
Since we expect $\varepsilon_\ell \sim g\,$, such terms are
irrelevant in {\gital v}\, up to $g^3$. In the sequel we
shall trust in the above arguments and omit the
${1\over 2}$--terms entirely.

Deleting the divergent pieces this way (in e.g. (\ref{3i})
and (\ref{3j})), $I$, $J$, $Y$ become well defined integrals
whose asymptotic series are known \cite{doja}$\,$: 
\begin{eqnarray} \label{3rei}
  I &=& {\pi^2 \over 90} - {\varepsilon^2 \over 24}
   + {\varepsilon^3 \over 12\pi} + {\varepsilon^4 \over 32 \pi^2}
   \ln \left( \varepsilon \right) - {c\, \varepsilon^4
   \over 64\pi^2} + O\left( \varepsilon^6\right)  \\  \label{3rej}
  J &=& {1\over 12}-{\varepsilon\over 4\pi} + \ldots \;
     = \; -{1\over \varepsilon} \partial_\varepsilon I \;
     = \; {1\over \varepsilon^2} L  \\ \label{3rey}
  Y &=& {1\over 8}-{\varepsilon\over 4\pi} + \ldots \; = \;
        {3\over \varepsilon^2}\left( I_0-I \right) \;\;
\end{eqnarray}
with $\varepsilon$ one of $\varepsilon_{t,\ell} = \beta
m_{t,\ell}$, $c={3\over 2} + 2 \ln (4\pi) - 2\gamma$ and
$\gamma$ the Euler constant. In the massless limit, the free
energy (\ref{3f}) is now recognized to be $n$ times that of
ordinary blackbody radiation.

The contributions to {\gital v}\, are now added up as
$F + \hbox{{\gital v}\,}_M + \hbox{{\gital v}\,}_{AAAA}$
and filled with details$\,$:
\begin{eqnarray}
  \hbox{{\gital v}\,} &=& nVT^4 \,\left[\, -2I_t
  - \varepsilon_t^2 J_t  + {1\over 2} \,\left[\, -2I_\ell -
  \varepsilon_\ell^2 J_\ell \,\right]\, + I_0\right. \nonumber \\
\label{3wr1}
  & & \left. \hspace{1.6cm} + \; {g^2 N \over 6} \left( 2J_t
   + J_\ell - Y_\ell + {\alpha\over 8} \right) \left( 2J_t
   + J_\ell + 2Y_\ell \right)  \,\right]\, \\  \label{3wr2}
  &=& {\rm const} + {nVT^4 \over 4\pi} \left(
   {\varepsilon_t^3\over 3} - \varepsilon_t \, g^2 N
   {5+\alpha\over 24} \;\, +\, {\varepsilon_\ell^3\over 6}
   - \varepsilon_\ell \, g^2 N {1+\alpha\over 16} \;
   + \ldots \; \right) \;\; . \qquad \quad
\end{eqnarray}
There it is, the announced wrong result$\,$: {\gital v}\,
depends on $\alpha$. Nevertheless, the structure is
appealing$\,$: the parameters $\varepsilon_t$ and
$\varepsilon_\ell$ do not mix, the only extremum is a minimum,
and its position has the right order $g^2N$ of magnitude. But,
apart from this, the minimum positions $\varepsilon_t^2 = g^2N
(5+\alpha )/24$ and $\varepsilon_\ell^2 = g^2N (1+\alpha )/8$
give no sense$\,$: {\sl which} $\alpha\, ? \,$ Including the
FP--term, with the means worked out in Sec.~V, does not help
out of this dilemma, because it only leads to minor changes.
To be specific, in (\ref{3wr2}) $5+\alpha$ becomes $6+\alpha$
and $1+\alpha$ turns into $(2+3\alpha )/3$.

%
%
   
\vspace{1.5cm} \centerline{\bf
IV. \ SCALAR ELECTRODYNAMICS } \vspace{1cm}  

For a first application of the ''even'' functional (\ref{2vvv}),
we appreciate scalar ED as a suitable example. Remember that
this system is an ideal toy model \cite{sed} to the gluon plasma,
with view to the identical diagram structure, the need of
resummation as well as to its physical gross features. The
Lagrangian, to be studied, is given by
\begin{equation} \label{4lbu}
   {\cal L} _\bullet = (D^\mu \phi )^\ast D_\mu \phi
   - {1\over 4} F^2 - {1\over 2\alpha_\bullet}
   \left(\partial A\right)^2 \quad
\end{equation}
with $D_\mu = \partial_\mu - igA_\mu$ and $F_{\mu\nu}
= \partial_\mu A_\nu - \partial_\nu A_\nu$. By again identifying
the fields (here$\,$: $\phi$ and $A$), the trial Lagrangian reads
\begin{equation} \label{4tri}
   {\cal L} = (\partial^\mu \phi)^\ast \partial_\mu \phi
   - {1\over 4} F^2 - {1\over 2\alpha} (\partial A)^2
   + {1\over 2} A(MA) - m_s^2 \phi^\ast \phi \quad . \;\;
\end{equation}
Its propagators are (\ref{2g}) and $1/\,\left[\, m_s^2-Q^2
\,\right]\,$ for photons and scalars, respectively. Here we
concentrate on the spectum of real excitations. Hence, the mass
matrix $M$ is that of the the $\lambda$--case (\ref{2lamb}).
The variational parameters in the above trial theory are
$\lambda_t$, $\lambda_\ell$ and the scalar mass $m_s$. The
Lagrangian (\ref{4tri}) turns into the effective Lagrangian
(at order $g^2$) of hot scalar ED \cite{sed} at the values
$\lambda_t = \lambda_\ell=1$ and $m_s^2=g^2T^2/4$. So, within
$O(g^2)$, the parameter space includes the exact answer (to be
derived by variation). Note that both, original and trial
theory, are invariant under regauging the photon field by
$\delta A = - \partial \Lambda$. By definition, the decoupling
ghost terms are kept apart from the above Lagrangians. But the
Faddeev--Popov compensation must be taken into account in the
partition function either by ghosts or as a determinant.

\vspace{1cm} \centerline{\bf
IV A. \ The ''even'' functional of scalar ED } \vspace{.4cm}

Recalling Sec.~II C, the partition function of scalar ED
may be written as
\begin{equation} \label{4zbu}
   Z_\bullet = {1\over Z_B^\bullet} \;{\rm det}^\prime
   (\beta^2\partial^2 ) \,{\cal N} \int {\cal D} \left\{ A_\mu
   \phi^\ast \phi \right\}\; e^{-S_{\bullet\bullet}}\quad . \;\;
\end{equation}
The prime on the Faddeev--Popov determinant excludes the
zero--eigenvalue (see also Appendix A). To specify ${\cal E}$
and ${\cal O}$ in (\ref{2vvv}), in the case at hand, we read off
from (\ref{4lbu}) that
\begin{equation} \label{4odd}
  {\cal O} = - \int^\beta {\cal L} _1 \;\quad \mbox{with}
  \;\quad {\cal L} _1 \; = \; igA^\mu \phi^\ast \partial_\mu
  \phi - ig (\partial_\mu \phi^\ast ) A^\mu \phi \;\; , \;\;
\end{equation}
while $ -\int^\beta {\cal L} _2$ with ${\cal L} _2 = g^2 A^\mu
A_\mu \phi^\ast \phi$ is part of ${\cal E}$ together with the
quadratic terms in (\ref{4lbu}). The index on ${\cal L}$ refers
to $g$--powers. With $\alpha = \alpha_\bullet$, as required for
(\ref{2vvv}) to be valid, we may thus write the ''even''
functional as
\begin{equation} \label{4vau}
  \hbox{{\gital v}\,} = F + T \left\langle \int^\beta
  {1\over 2} A(MA) - m_s^2 \int^\beta \phi^\ast\phi - \int^\beta
  {\cal L} _2 - \ln \,\left[\, {\rm ch} \left( \int^\beta
  {\cal L} _1 \right) \,\right]\, \right\rangle \;\; , \;\;
\end{equation}
where $F=-T\ln(Z)$, and the trial partition function $Z$ is
given by (\ref{4zbu}) with all bullets stripped off there.

There is a high (but probably inevitable) price to be paid for
the physical consistency reached with the above formulation$\,$:
it obviously contains fields in arbitrary high powers (instead
of only quartic). For the explicit evaluation of (\ref{4vau})
one is, apparently, forced again into a perturbative expansion,
namely that of the logarithmic term (\, $\ln\,\left[\, {\rm ch}
(x) \,\right]\, = x^2/2-x^4/12+x^6/45-\ldots$\,). But note, at
least, that this expansion looks much simpler than diagrammatic
thermodynamics$\,$: here the seagull vertex does not occur
in higher powers. If, by any reason, terms of order $g^4$ may
be neglected, then the functional simplifies to
\begin{equation} \label{4trunc}
  \hbox{{\gital v}\,}_{\rm trunc} = F + T \left\langle
  \,\int^\beta {1\over 2} A(MA) - m_s^2 \int^\beta \phi^\ast\phi
  - \int^\beta {\cal L} _2  - {1\over 2} \left( \int^\beta
  {\cal L} _1 \right)^{\! 2} \,\right\rangle \;\; . \;\;
\end{equation}
In the following, while demonstrating the value of
(\ref{4vau}), we shall in fact restrict to the truncated
version (\ref{4trunc}).

\vspace{1cm} \centerline{\bf
IV B. \ Evaluation of {\gital v}\,$_{\rm trunc}$ } \vspace{.4cm}

Let us group the above five contributions into ''bare'' and
interaction terms$\,$:
\begin{equation} \label{4vsum}
  \hbox{{\gital v}\,}_{\rm trunc}
  = \hbox{{\gital v}\,}_0 + \hbox{{\gital v}\,}_{\rm int}
    \quad \mbox{with} \quad
  \hbox{{\gital v}\,}_0 = F + \hbox{{\gital v}\,}_M
  + \hbox{{\gital v}\,}_{m_s^2} \;\; , \;\;
  \hbox{{\gital v}\,}_{\rm int}
  = \hbox{{\gital v}\,}_{AA\phi\phi}
  + \hbox{{\gital v}\,}_{\rm square} \;\; . \;\;
\end{equation}
For {\gital v}\,$_0$ we are well prepared from Sec.~III$\,$:
strip off the colour factor $n$ from (\ref{3vm}) and
(\ref{3f}). Of course, with view to (\ref{aend}), the free
energy $VT^4 (-2 I_s )$ of the scalars has to be added now. As
the scalars have constant mass, $I_s$ is given by (\ref{3i})
with index $\ell$ replaced by $s$. For
{\gital v}\,$_{m_s^2}$ note, that the average
$\left\langle \phi^\ast \phi \right\rangle$ equals
$\sum_P S(P)$ with the scalar propagator given by
$S(P)=-1/(P^2-m_s^2) \equiv - \Delta_s $. Thus, in particular
\begin{equation} \label{4vm} \hspace{-.2cm}
  \hbox{{\gital v}\,}_{m_s^2} = - V m_s^2 \left\langle
  \phi^* \phi \right\rangle = - V T^4 L_s \;\; , \;\;
  L_s = \varepsilon_s^2 J_s \;\; , \;\; \varepsilon_s \equiv
  \beta m_s \;\; , \;\; J_s = -\beta^2 \sum \Delta_s \;\; , \;\;
\end{equation}
and in total
\begin{equation} \label{4vtot}
  \hbox{{\gital v}\,}_0 = V T^4 \left( - 2I_t - L_t - I_\ell
  - {1\over 2} L_\ell + I_0 - 2 I_s - L_s \right) \;\; . \;\;
\end{equation}

Among the interaction terms, one is pretty simple$\,$:
\begin{equation} \label{4v2}
   \hbox{{\gital v}\,}_{AA\phi\phi} = - V g^2 \left\langle
   \phi^*\phi \right\rangle \left\langle A^\mu A_\mu
   \right\rangle = VT^4 \, g^2 \left( 2 J_t + J_\ell
    \right) J_s  + VT^4 \, g^2 \alpha J_0 J_s \;\;\;
  = \; F_\bullet^{\circ\!\circ} \;\; . \;\;
\end{equation}
Of course, $J_0=1/12$ even in the $\lambda$--case. To the right
in (\ref{4v2}), we have noted, that
{\gital v}\,$_{AA\phi\phi}$ precisely
equals the perturbative free energy contribution from the
diagram $\circ\!\circ$ (one loop scalar, one photonic). But
here the lines represent massive propagators, making
$\circ\!\circ$ depending on variational parameters. One may
speculate, that the remaining term {\gital v}\,$_{\rm square}$
could correspond to the diagram $\ominus$ (the inner line
photonic). This is indeed the case, see (\ref{4vs}) below.
Classes of diagrams were whisked away in the treatment of
Sec.~III.

The first steps in treating {\gital v}\,$_{\rm square}$ are
straightforward$\,$: Fourier transform all fields, Wick
decompose (the $\circ\!\! - \!\!\circ$--diagram drops out due
to odd summand) and use (\ref{2aa}). One obtains
\begin{equation} \label{4vs}
  \hbox{{\gital v}\,}_{\rm square} = - V {1\over 2} g^2
  \sum_Q \sum_P \, (2P-Q)^\mu\, G_{\mu\nu} (Q)\, (2P-Q)^\nu\,
  \Delta_s \Delta_s^ - \;\;\; = \; F_\bullet^\ominus \;\;
\end{equation}
with $\Delta_s^- = 1 / \,\left[\, (Q-P)^2 - m_s^2 \,\right]\,$.
There is gauge--fixing dependence in {\gital v}\,$_{\rm square}$
through the propagator $G$ from (\ref{2g}). After some algebra
one obtains that
\begin{equation} \label{4valph}
   \hbox{{\gital v}\,}_{\rm square} -
   \hbox{{\gital v}\,}_{\rm square}^{\quad (\alpha =0)}
   \;\; = \; - V T^4 \, g^2 \alpha J_0 J_s \;\; , \;\;
\end{equation}
which cancels the $\alpha$--term of (\ref{4v2}). Thus, in the
case of scalar ED and in its truncated ''even'' functional,
there is no gauge--fixing dependence. This is true for all values
of our variational parameters and for any mass matrix version.

In (\ref{4vs}) at $\alpha =0$ terms with $Q^\mu$ or $Q^\nu$ are
projected out. The remaining sandwiches are $P\hbox{\logo A} P =
- \,\left[\, {\bf p}^2 - ({\bf p} {\bf q} )^2 / q^2 \,\right]\,$
and $P{\sf B} P = P^2 - (PQ)^2 / Q^2 - P\hbox{\logo A} P$.
Expecting the structure (\ref{3wr2}), at least the terms linear
in $\lambda_t$, $\lambda_\ell$, $m_s$ must be detailed. We
therefore form differences as e.g. $\Delta_{t\ell} \equiv
\Delta_t - \Delta_\ell$, add (\ref{4vs}) to (\ref{4v2}), write
$J_{t\, 0} \equiv J_t-J_0$ etc. and split {\gital v}\,$_{\rm int}$
into a constant ($v_0$), terms linear in such differences ($v_1$)
and the rest ($v_2$), which is certainly of higher order. Then,
some terms of $v_1$ (easily identified in (\ref{4vv2}) below)
are regrouped into $v_2$, because they are of higher order by
other reasons. Note that $\sum_P \left( 4P^2Q^2 - 4(PQ)^2 + Q^4
\right) / \,\left[\, Q^2 P^2 (Q-P)^2 \,\right]\, = 2\sum_P 1/P^2$.
We obtain$\,$:
\begin{eqnarray} \label{4vint}   \hspace{-.3cm}
  \hbox{{\gital v}\,}_{\rm int}
 &=& V g^2 \left( v_0 + v_1 + v_2 \right)
     \qquad \mbox{with} \\  \label{4vv0}
 v_0 &=& \sum_Q \sum_P {1\over Q^2 P^2} \left( 3 - 2 {Q^2 P^2
     -(QP)^2 \over Q^2 (Q-P)^2 } \right)\; =\;
     {5\, T^4 \over 288 } \;\; , \\ \label{4vv1}
 v_1 &=& T^4 J_0 \left( 2 J_{t\, 0} + 3 J_{s\, 0} \right) + 2
     \sum_Q \Delta_{t\,\ell}(Q)\sum_P {1\over P^2 (Q-P)^2 }
     \,\left[\,{\bf p}^2 - {({\bf p} {\bf q} )^2 \over q^2 }
     \,\right]\, \;\; , \\
 v_2 &=& \sum_Q \sum_P \left( 2 \Delta_{t\, 0} \Delta_{s\, 0}
     + 2 \,\left[\, {\bf p} ^2 - {({\bf p} {\bf q} )^2 \over
     q^2 } \,\right]\, \Delta_{t\,\ell} \left( \Delta_s
     \Delta_s^- - \Delta_0 \Delta_0^- \right) + \Delta_{\ell\, 0}
     \Delta_{s\, 0}\right.\nonumber\\ \label{4vv2}
  & & \left. -\; 2\,\left[\, P^2 - {(PQ)^2 \over Q^2} \,\right]
     \,\Delta_\ell \left( \Delta_s \Delta_s^- - \Delta_0
     \Delta_0^- \right) + {1\over 2} \Delta_{\ell\, 0} Q^2
     \Delta_0 \Delta_0^- \right)  \;\; . \;\;
\end{eqnarray}
In each term of (\ref{4vv2}) the argument of the first propagator
is $Q$, and $P$ that of the second. To understand why the
last two terms of (\ref{4vv2}) are less than $O(g)$, the first
one can be rewritten as $-2m_s^2 \sum_{Q,P} \Delta_\ell \Delta_s
\Delta_s^-$ plus some products of differences. But for the last
term in (\ref{4vv2}) only a detailed analysis (of the type done
in Appendix B) reveals its order $T^4 g^2 \ln (g) $ of magnitude.
Such terms are known to occur in the perturbation expansion of
the free energy \cite{arzk}.

Up to order $g^3$ in {\gital v}\, (or order $g$ in $v_1$) only 
the line (\ref{4vv1}) needs further study. Note at first that, 
formally, the expression (\ref{2pi}) (at $N=1$ here) appears in 
this line. So, the machinery ''knows'' of the leading--order
longitudinal polarization function. There are two ways to
evaluate $\sum_Q \Delta_{\ell\, t} {\mit \Pi}_\ell$ (for
later use, we detail both). First, one may cancel
${\mit \Pi}$--functions with those in the trial propagators
(using ${\mit \Pi}_\ell=3m^2-2{\mit \Pi}_t$), write
$v_1$ in terms of basic integrals as
\begin{equation} \label{4eck}
  \sum_Q \Delta_{\ell\, t} {\mit \Pi}_\ell = 
     - T^4 \left( \, {2 \over \lambda_t^2 } L_t 
     + {1 \over \lambda_\ell^2} L_\ell - {g^2 \over 3} J_t 
     \,\right)\,  \;\; 
\end{equation}
and proceed with expanding the latter (see below). Note that
these cancellations are possible in the $\lambda$--case only.
The alternative second way is by far the easier and more
enlightening one. As is basic to the dimensional reduction
method \cite{kaj,bran,ay} and to various related thermodynamic
calculations (e.g. \cite{bumu,arzk}), a frequency sum may be
occasionally reduced to its $Q_0=0$ term. This step, if valid
\cite{sed,frosch}, rests on the structure of a massive
propagator $\Delta (Q)$ and usually prepares its soft part
while contributions from non--zero hard frequencies are of
higher order. Of course, there must be no hard part in a sum
under such study. In fact, (\ref{4eck}) is an ideal example
for the above. Moreover, at $Q_0=0$ the polarization functions
${\mit \Pi}_\ell$ and ${\mit \Pi}_t$ reduce to constants,
namely $3m^2$ and $0$, respectively$\,$:
\begin{equation} \label{4null}
  \sum_Q \Delta_{\ell\, t} {\mit \Pi}_\ell = 
     3 m^2 {T \over 2 \pi^2} \int_0^\infty \! dq \, q^2
     \left( {1\over q^2} - {1 \over q^2 + \lambda_\ell^2 3
     m^2 } \right) \; = \; {T^4 g^2 \over 12 \pi }
     \left( {g \lambda_\ell \over 
\sqrt{3\,}^{\hbox to0.2pt{\hss$
\vrule height 2pt width 0.6pt depth 0pt $}\;\!} 
     } \right)  \;\; . \;\;
\end{equation}
Dependence on $\lambda_t$ has dropped out, with the reason
readily detected in the vanishing factor ${\mit \Pi}_t$ of
$\lambda_t^2$, i.e. in the absence of a (squared) magnetic
mass at order $g^2$. In the same manner the soft parts of
$J$--integrals are obtained$\,$:
\begin{equation} \label{4jot}
   J_\ell = J_0 - {1\over 4\pi} \,{g\lambda_\ell \over
\sqrt{3\,}^{\hbox to0.2pt{\hss$
\vrule height 2pt width 0.6pt depth 0pt $}\;\!} }\; + \;\ldots
    \qquad , \qquad J_t = J_0 + 0\; + \;\ldots \;\;
\end{equation}
with again no dependence on $\lambda_t$ by the same reason.

To complete the evaluation, note that the scalar contributions
$J_s$, $L_s$, $I_s$ are $m$--case objects, hence their
expansions are given by (\ref{3rei}) and (\ref{3rej}) with 
$\varepsilon = \varepsilon_s = \beta m_s$. For the remaining
$\lambda$--case integrals in {\gital v}\,$_0$, (\ref{4vtot}), 
apparently, the sums $L_t$ and $L_\ell$ are still to be studied
separately. For this somewhat delicate task see Appendix B.
As a result, both $L$ start with a $g^2\lambda^2$--term whose 
prefactor $\kappa$ diverges logarithmically. From (\ref{b3})
to (\ref{b5})$\,$:
\begin{equation} \label{4lex}
  L_\ell = 2\,\kappa \, {g^2 \lambda_\ell^2 \over 3 }
  - {1\over 4\pi}  \left( { g \lambda_\ell \over
\sqrt{3\,}^{\hbox to0.2pt{\hss$
\vrule height 2pt width 0.6pt depth 0pt $}\;\!}
  } \right)^3\; + \; \ldots \quad , \quad
  L_t = \left( {1\over 24} - \kappa \right) {g^2
  \lambda_t^2 \over 3 }\; + 0\; + \;\ldots \;\; . \;\;
\end{equation}
Now note that the singular piece $\kappa$ drops out in
{\gital v}\,$_0$, because there $L$ appears in the combination
$2I + L = 2I_0 + \left\{ 1 - 2\int_0^\lambda\! d\lambda\,
{1\over \lambda} \right\} L$, see (\ref{3ilrel}),
and the curly bracket is a projector$\,$: 
$\left\{ \;\right\} \lambda^2 = 0$. One might ask for the fate
of the singular terms in (\ref{4eck}). They drop out there
by cancellation, and (\ref{4null}) derives again.

\vspace{1cm} \centerline{\bf
IV C \ Minimizing {\gital v}\,$_{\rm trunc}$ at order $g^3$ }
       \vspace{.4cm}

In the preceding subsection, the expansions were driven just as
far as to allow for writing down the functional {\gital v}\,
up to third order in the coupling $g$. Of course, as in Sec.~III,
we anticipate that the solutions to $m_s$ and $\lambda_{t,\ell}$
will be $O(gT)$ and $O(1)$ in magnitude, respectively. By
combining the details of the preceding subsection one obtains
\begin{equation} \label{43v}
  \hbox{{\gital v}\,}^{\;{\rm to}\; g^3} = V\, T^4\,
  \left( - 2\, {\pi^2 \over 45} + {5\, g^2 \over 288}
  + {g^3 \over 24\pi
\sqrt{3\,}^{ \hbox to0.2pt{\hss$
\vrule height 2pt width 0.6pt depth 0pt $}\;\!}
  } \,\left[\, {\lambda_\ell^3 \over 3} - \lambda_\ell
  \,\right]\, + {1 \over 12\pi} \left[ \, (\beta m_s)^3  -
  {3 g^2 \over 4} \beta m_s \,\right] \,\right) \;\; . \;\;
\end{equation}
It still has the structure of (\ref{3wr2}). The variational
parameters do not couple, which is specific to the order
considered. The absence of any dependence on
$\lambda_t$ was already understood, although merely technically
(see also Sec.~VI). The above {\gital v}\,, when plotted over
the $\lambda_\ell$--$\lambda_t$--plane, has the form of a long
gutter. The resolution of this defect is deferred to 
Sec.~IV D.

Minimizing (\ref{43v}) with respect to $m_s$ and
$\lambda_\ell$ gives the values
\begin{equation} \label{43val}
  m_s^{\rm min} \, = \;{1\over 2}\, g \, T \qquad ,
  \qquad \lambda_\ell^{\rm min} \, = \; 1 \quad , \quad
\end{equation}
as expected. We immediately look for the value of the above
{\gital v}\, taken at these parameters, which
is the height of the bottom of the gutter$\,$:
\begin{equation} \label{43min}
  \hbox{{\gital v}\,}^{\;\rm min} = V\, T^4\, \left( - 2\,
  {\pi^2 \over 45} + {5\, g^2 \over 288} - {g^3 \over 12\pi}\,
  \,\left[\, {1\over 3
\sqrt{3\,}^{\hbox to0.2pt{\hss$
\vrule height 2pt width 0.6pt depth 0pt $}\;\!}
   } + {1\over 4} \,\right]\, \right) \;\;
\end{equation}
with the last term in the square bracket being due to the
scalars. The minimum perfectly agrees with the perturbative
free energy up to $g^3$. The $g^3$--term, the correlation
energy, was given by Kalashnikov and Klimov \cite{kakl}
(eq.~(19) there, taken at $\lambda =\mu=0$ and $e=g$).
In summary, for scalar electrodynamics and up to 
the third $g$ power, the ''even'' variational functional
has all required properties, namely gauge fixing
independence, the right minimal value and (apart from
degeneracy) the right minimum position.

\vspace{1cm} \centerline{\bf
IV D. \ Solution to the gutter problem } \vspace{.4cm}

The missing dependence on $\lambda_t$ in (\ref{43v}) is, as
already noticed, an artifact of the restriction to order $g^3$
of the functional. The problem merely is how to go one order
higher {\sl within} the expansions so far developed. First of
all, we notice that $g^4$--terms are allowed within the truncated
functional, although the neglected next term of $\ln \,\left[\,
{\rm ch} (x) \,\right]\,$ does contribute at order $g^4$ too.
However, the latter is a constant at this order; variational
parameters appear at $g^5$.

Let us try to avoid expansions, and let the collection of all
terms containing $\lambda_t$ be denoted by {\gital v}\,$_{\rm
trunc}^{\,(t)}$. Up to an additive constant, it may be written
as
\begin{equation} \label{44w}  \hspace{-.2cm}
   \hbox{{\gital v}\,}^{\,(t)}_{\rm trunc} =
   V\, T^4\, {\cal U}_t + V\, g^2\, v_2^{(t)} \;\quad
   \mbox{with} \quad {\cal U}_t = - L_t + 2\int_0^{\lambda_t}\!
   d\lambda \, {1\over \lambda}  L_t (\lambda_t=\lambda )
   + {1\over \lambda_t^2 } L_t \;\; .\;\;
\end{equation}
Here, $v_2^{(t)}$ is made up of the first two terms in
(\ref{4vv2}), but leave $v_2^{(t)}$ aside for a moment. Then,
the minimum condition may be given the form of a product
\begin{equation} \label{44dw}
  0 = \partial_{\lambda_t} {\cal U}_t
  = \left( {2\over \lambda_t} L_t - \partial_{\lambda_t} L_t
  \right) \cdot \left( 1 - {1 \over \lambda_t^2 } \right) \;\;
\end{equation}
with the first factor ''unknown'', but the second reaching zero
at $\lambda_t^2 = 1$ as desired. To be sure that this zero
corresponds to a minimum, the first factor must be shown to be
positive. We shall do so at the end of Appendix B. There, the
first factor is also seen to be of order $g^4$ and to vary as
$\lambda_t^3$ for small $g$, see (\ref{b9}). Hence, ${\cal U}_t$
has a Higgs--type shape ${\cal U}_t \sim {\rm const} - g^4
\lambda_t^2 + {1\over 2}g^4 \lambda_t^4 + \ldots$ with a maximum
at the origin. The curvature of the gutter sets in one order
higher, indeed. A plot of {\gital v}\, now merely looks
like a long bath--tub.

The above construction only works if the correction $v_2^{(t)}$
remains below the order $O(g^2)$. Its first term is the first
in (\ref{4vv2}) and is of order $T^4g(J_t-J_0)$ in magnitude.
With view to (\ref{4jot}) it is indeed below $g^2$. For the
second contribution (the second in (\ref{4vv2}) but with
$\Delta_{t\, 0}$ in place of $\Delta_{t\,\ell}$) we need a bit
of calculation. Both sums may be considered ''soft'', i.e.
$n(x) \to T/x\,$ is allowed, thereby preparing the contribution
of interest. All propagators are represented spectrally. For
the two frequency sums, eq. (6.6) of \cite{nt} is used repeatedly.
The result is a 3--momentum double integral over (among other
factors)  $\int\! dx {1\over x} \left( \rho_t^{(\lambda )}
(x,q)-\rho^{(0)} (x,q) \right)$. But, due to the sum rule
(\ref{csumt}), this factor vanishes, q.e.d.

%
%

\vspace{1.5cm} \centerline{\bf
V. \ YANG--MILLS FIELDS \ \ ($\;$the gluon plasma$\;$)}
     \vspace{1cm}
   
For treating the non--Abelian theory (\ref{2lbu}) in its ''even
version'', we use Sec.~IV as a guideline. Hence, first of all,
we strip off the ghost terms from ${\cal L}_\bullet$, ${\cal L}$
and introduce the index ''no'' for such reduced Lagrangians$\,$:
\begin{eqnarray} \label{5sbu}
  S_\bullet^{\,\rm no} &=& - \int^\beta {\cal L}_\bullet^{\,\rm
  no}  = - \int^\beta \left( {\cal L}_0 + {\cal L}_1
  + {\cal L}_2 \right)  \qquad \mbox{with}  \\ \label{5l12}
  {\cal L}_1 &=& - g \left( \partial_\mu A_\nu^a \right)
  f^{abc} A^{\mu\, b}  A^{\nu\, c}  \;\; , \;\;
  {\cal L}_2 = - {1\over 4} g^2 f^{abc}f^{ars} A_\mu^b A_\nu^c
  A^{\mu\, r} A^{\nu\, s} \;\; . \;\;
\end{eqnarray}
Here, ${\cal L}_0$ is the quadratic part of
${\cal L}_\bullet^{\,\rm no}$, hence including the gauge
fixing$\,$: ${\cal L}_0 = - (F^a)^2/4 - (\partial A^a)^2 /
(2\alpha_\bullet)\,$. The Faddeev--Popov determinant now
depends on the gauge field and is thus subject to functional
integrations. But for convenience we may split off its bare
factor. The partition function, still waiting for its
even--odd decomposition, so far reads
\begin{equation} \label{5zbu}
  Z_\bullet = {1\over Z_B^\bullet} {\rm det}^\prime
  \left(\beta^2 \partial^2 \delta^{ab} \right) {\cal N} \int
  {\cal D} A_\mu^a e^{-S_\bullet^{\,\rm no} - S_{FP}} \;\; .\;\;
\end{equation}
The two factors in (\ref{5zbu}), which obviously stand for the
FP--determinant ${\rm det}^\prime \left( \beta^2\partial D
\right)$, derive through
\begin{eqnarray}
  {\rm det}^\prime \left( \beta^2 \partial D \right)
  &=& {\rm det}^\prime \left( \beta^2 \partial^2 \delta^{ab}
    \right) {\rm det}^\prime \left( \,\left[\,
    \partial^2\delta^{ab} - \partial^\mu g f^{abc} A_\mu^c
    \,\right]\, {1\over \partial^2\delta^{ab}} \right)
    \nonumber \\ \label{5det}
 &\equiv&  {\rm det}^\prime \left( \beta^2 \partial^2
   \delta^{ab} \right) {\rm det}^\prime  \left( 1 + {\cal W}
   \right) \;\equiv\;{\rm det}^\prime \left(\beta^2 \partial^2
   \delta^{ab} \right)\, e^{-S_{FP}}\;\; ,\;\;
\end{eqnarray}
where by ${\cal W}$ the part odd in the gauge field is
prepared$\,$:
\begin{equation} \label{5w}
   {\cal W} = - g f^{abc} \partial^\mu  A_\mu^c {1\over
   \partial^2} \;\; . \;\;
\end{equation}
The first $\partial^\mu$ acts on $A_\mu^c$ {\sl and} all
functions that follow. We read off from (\ref{5det}) that
\begin{eqnarray} \label{5sfp}
   S_{FP} = - \ln \,\left[\, {\rm det}^\prime \left( 1 +
   {\cal W} \right) \,\right]\,
  &=& - \, {\rm Tr}\,^\prime  \ln \left( 1 + {\cal W} \right)
    \nonumber \\
  &=& - {1\over 2} \, {\rm Tr}\,^\prime \,\ln \left( 1 -
  {\cal W}^2 \right) - {1\over 2} \, {\rm Tr}\,^\prime \,\ln
  \left( { 1 + {\cal W} \over 1 - {\cal W} }\right)\;\; .\quad
\end{eqnarray}
In the second line, clearly, the even--odd decomposition is
achieved. But the second equality in (\ref{5sfp}) (first line)
is delicate, because all eigenvalues of $1+{\cal W}$ have to
be positive, but are not. While this point needs care in
exactly solvable models \cite{wipf}, here we may be content
with a crude argument. For the intended comparison with
perturbation theory, the above logarithms are expanded
anyways. Hence (\ref{5sfp}) is merely a formal compact
notation for series to be generated \cite{direu}.

We are ready to form the non--Abelian ''even'' action
$S_{\bullet\bullet}$ through $\,S_\bullet^{\,\rm no} + S_{FP}
\to S_{\bullet\bullet} = {\cal E} - \ln \,\left[\, {\rm ch}
\left( {\cal O} \right) \,\right]\,\,$ with
${\cal E}$, ${\cal O}$ given by
\begin{eqnarray} \label{5even}
  {\cal E} &=& -\int^\beta \left({\cal L}_0 + {\cal L}_2 \right)
  - {1\over 2}\, {\rm Tr}\,^\prime \,\ln \left( 1
  - {\cal W}^2 \right) \;\; ,  \\ \label{5odd}
 {\cal O} &=& - \int^\beta {\cal L}_1 - {1\over 2} \,
   {\rm Tr}\,^\prime  \,\ln \left({ 1+{\cal W} \over 1
   -{\cal W}} \right) \;\; . \;\;
\end{eqnarray}
The trial theory has remained unchanged. It is that of 
Sec.~III. The trial partition function is given by (\ref{5zbu})
without the bullets, at $S_{FP} = 0$ and with $S^{\rm no}
= - \int^\beta {\cal L}^{\rm no}$. The free energy $F$ is
(\ref{3f}). Thus, the ''even'' functional (\ref{2vvv}) of the
gluon system (taken at $\alpha = \alpha_\bullet$) reads
\begin{eqnarray}
  \hbox{{\gital v}\,} &=& F + T \left\langle \int^\beta
  {1\over 2} A^a (MA^a) - \int^\beta {\cal L}_2 - {1\over 2}
  \, {\rm Tr}\,^\prime \,\ln \left( 1 - {\cal W}^2 \right)
  \right. \nonumber \\  \label{5vau}
 & & \left. \hspace{1.5cm} - \;\ln \,\left[\, {\rm ch}
  \left( \int^\beta {\cal L}_1 + {1\over 2} \, {\rm Tr}\,^\prime
  \,\ln \,\left[\, {1 + {\cal W} \over 1 - {\cal W} } \,\right]\,
  \right) \,\right]\, \right\rangle  \;\; . \quad
\end{eqnarray}
The first two terms form the bare part
{\gital v}\,$_0$ and are familar from Sections III,IV$\,$:
\begin{equation} \label{5v0}
   \hbox{{\gital v}\,}_0 = n V T^4 \left( - 2 I_t - L_t
   - I_\ell - {1\over 2} L_\ell + I_0 \right) \;\; . \;\;
\end{equation}

As in Sec.~IV, we expand the logarithms up to ${\cal W}^2$ to
reach a reasonable simple ''truncated version''. Since
$\, {\rm Tr}\,^\prime \,{\cal W} = 0\,$, no such term arises
from the last logarithm. Thus,
\begin{equation} \label{5trunc}
   \hbox{{\gital v}\,}_{\rm trunc} = \hbox{{\gital v}\,}_0
   + \hbox{{\gital v}\,}_{\rm int} \quad , \quad
   \hbox{{\gital v}\,}_{\rm int} = \hbox{{\gital v}\,}_{AAAA}
   + \hbox{{\gital v}\,}_{FP} +
   \hbox{{\gital v}\,}_{\rm square} \quad
\end{equation}
with
\begin{equation} \label{5fpsq}
   \hbox{{\gital v}\,}_{FP}
   = {T\over 2}\left\langle \, {\rm Tr}\,^\prime \,{\cal W}^2
   \right\rangle \quad , \quad \hbox{{\gital v}\,}_{\rm square}
   = - {T\over 2} \left\langle \left(
 \int^\beta {\cal L}_1 \right)^{\! 2}\,\right\rangle \;\; .\;\;
\end{equation}
The contribution {\gital v}\,$_{AAAA}$ is given by
(\ref{3aaaa}). It agrees with the perturbative free energy
contribution from the tadpole diagram (both lines gluons)$\,$:
{\gital v}\,$_{AAAA} = F_\bullet^{\circ\!\circ}\,$. Compared to
Sec.~III, there are two additional terms in (\ref{5trunc})$\,$:
the last two. By analogy with Sec.~IV we expect that they equal
the two other diagrams at second order, which were missing in
Sec.~III. Indeed, taking the trace of ${\cal W}^2$ with states
$(\beta V)^{-1/2} e^{-iPx}$, using $f^{abc}f^{abc}=Nn$ and
through Wick decomposition, we obtain
\begin{eqnarray} \label{5vfp}
 \hbox{{\gital v}\,}_{FP} \;\, &=& \;\, nV {g^2 N\over 2}
   \sum_Q \sum_P G_{\mu\nu} (Q) {P^\mu (P-Q)^\nu \over P^2
   (Q-P)^2 }  \;\;\; = \; F_\bullet^\div \;\; , \;\; \\
 \hbox{{\gital v}\,}_{\rm square} &=& nV {g^2 N\over 2}
   \sum_Q \sum_P \,\left[\, (Q+P)^\lambda  G_{\lambda \nu}
   (Q-P) G^{\nu \rho} (Q)
   \hspace{4cm} \right. \nonumber \\ \label{5vsq}
 & & \left.  - \; G_{\lambda \tau} (Q) G^{\lambda \tau} (Q-P)
    \, Q^\rho \,\right]\,  G_{\rho \mu} (P) (2Q-P)^\mu
    \;\;\; = \; F_\bullet^\ominus \;\; , \;\;
\end{eqnarray}
where, in (\ref{5vfp}) the symbol ''$\div$'' (with two out of
many dots) stands for the ghost loop with an inner gluon line.

Quite different from scalar ED, the gauge--fixing dependence
does not cancel in a manner independent of variational
parameters. Splitting the Greens function as $G = \chi + \alpha
{\sf D} \Delta_0$, we see that $\alpha$ occurs up to the third
power. The term $\alpha^3$ is contained in {\gital v}\,$_{\rm
square}$ only, and its prefactor vanishes. Collecting
$\alpha^2$-- and $\alpha$--terms one obtains
\begin{eqnarray} \label{5aq} \hspace{-.3cm}
   \hbox{{\gital v}\,}^{\, (\alpha^2)}\!  &=& - nV {g^2 N
   \over 4} \, \alpha^2 \,\sum_Q \sum_P {Q^4 \over P^4 (Q-P)^4 }
   P^\mu \chi_{\mu\nu} (Q) P^\nu  \; - \;\mbox{\footnotesize
   (the same at zero mass)} \;\; , \;\;  \\  \hspace{-.3cm}
 \hbox{{\gital v}\,}^{\, (\alpha )}\! &=& n V {g^2 N
   \over 2} \, \alpha \, \sum_Q \sum_P {1\over (Q-P)^4 }\,
   \left[\, \chi_\mu^{\;\;\mu}(Q) (Q-P)^2 - P^\mu \chi_{\mu\nu}
   (Q) P^\nu \hspace{2.4cm} \right.  \nonumber \\ \label{5a}
 & & \left. + \; Q^2 \left( P^2-Q^2 \right) \chi_{\mu\nu} (Q)
   \chi^{\mu\nu} (P) \,\right]\, \;
   - \;\mbox{\footnotesize (the same at zero mass)} \;\; . \;\;
\end{eqnarray}
The fact that (\ref{5aq}) and (\ref{5a}) vanish at zero mass
reflects gauge invariance of thermodynamic perturbation theory
at order $g^2$. For the next step, namely analysing
{\gital v}\,$_{\rm trunc}$ at order $g^3$, we need more$\,$:
(\ref{5aq}) and (\ref{5a}) must remain below $g^3$. This is the
case, as one may check e.g. by power counting. Remember that
perturbatively a $g^3$ only arises by dressing the
$g^2$--diagrams, whereby gauge invariance persists.

The strategy of further evaluation is now that of Sec.~IV, as
detailed above (\ref{4vint}). Since they are of higher order,
we temporarily omit the two $\alpha$--dependent terms (\ref{5aq})
and (\ref{5a}). In {\gital v}\,$_{\rm int}$ this
amounts to the replacement $G \to \chi = \hbox{\logo A}
\Delta_t + {\sf B} \Delta_\ell$. Then the terms ($v_1$) linear
in $\chi -\chi_0$ are isolated, and terms of higher order
-- others than in Sec.~IV -- move to $v_2$. But evaluation of $v_1$
runs through the steps in Sec.~IV B and, surprisingly,
ends up with the {\,\sl same\,} result as in Sec.~IV, namely
(\ref{4eck}) at $m_s=0$. Just to show prefactors$\,$:
\begin{equation} \label{5vint}  \hspace{-.2cm}
   \hbox{{\gital v}\,}_{\rm int}
   = n V g^2 N \left( v_0 + v_1 + v_2 \right) \; , \; v_0
   = {T^4 \over 144} \; , \; v_1 = {T^4 \over g^2 N
   \lambda_t^2} L_t + {T^4 \over 2 g^2 N \lambda_\ell^2}
   L_\ell - {T^4 \over 6} J_0  \; . \;
\end{equation}
The complete functional up to order $g^3$ (add
(\ref{5vint}) to (\ref{5v0})), does not depend
on $\lambda_t$ (gutter form) and reads
\begin{equation} \label{5uu}
  \hbox{{\gital v}\,}^{\;{\rm to}\; g^3}
  = \, {\rm const}\, + \, n V T^4 {1\over 2}
  \,{\cal U}_\ell \;\; 
\end{equation}
with the function ${\cal U}_{\ell}$ defined as
${\cal U}_t$ in (\ref{44w}) by changing the index.
Minimization gives $\lambda_\ell = 1$, as desired.
For the height of the minimum to order $g^3$ we
obtain
\begin{equation} \label{5min}
  \hbox{{\gital v}\,}^{\;\rm min} = n V T^4 \,\left[\,
  - {\pi^2 \over 45} + {g^2 N \over 144}  - {1\over 12\pi }
  \left( {g
\sqrt{N\,}^{\hbox to0.2pt{\hss$
\vrule height 2pt width 0.6pt depth 0pt $}\;\!}
   \over
\sqrt{3\,}^{\hbox to0.2pt{\hss$
\vrule height 2pt width 0.6pt depth 0pt $}\;\!}
   } \right)^3 \; \,\right]\, \;\; . \;\;
\end{equation}
This is equation (8.47) in \cite{kapu}. At $N=1$, the
correlation energy ($g^3$--term) agrees with the photonic 
one in scalar ED, see (\ref{43min}).

As in the Abelian case (Sec.~IV D) the functional
is expected to become convex with respect to $\lambda_t$
by including $g^4$--terms. However, at this point we run 
into non--Abelian difficulties. There are four terms to be
included. The first one is ${\cal U}_t$ (replace
${\cal U}_\ell$ in (\ref{5uu}) by ${\cal U}_\ell + 2\,
{\cal U}_t$), which has a minimum at $\lambda_t=1$. The
second term arizes from $v_2$ in (\ref{5vint}), a rather
lengthy expression (seven lines say) and so far not
evaluated. The third and fourth terms are the
$\alpha$--dependent pieces (\ref{5aq}), (\ref{5a}) and
cause the trouble. They should be (but are not) either
constant, or minimal at $\lambda_t=1$, too, or of lower
order in magnitude. Consider e.g.
the $\Delta_{t\, 0}$--part of the $\alpha^2$--term
(\ref{5aq}). If evaluated ''soft'' it vanishes (in the
manner noted at the end of Sec.~IV). At first glance,
as no UV--cutoff is needed, one might conclude that
{\gital v}\,$^{(\alpha^2)}_t = 0$ at all. However, it
appears that there is still a hard contribution, which in
turn needs no IR--cutoff. Because this is perhaps somewhat
unusual, let us state the result$\,$:
\begin{eqnarray}    \hspace{-.2cm}
   \hbox{{\gital v}\,}^{\,(\alpha^2)}_t
 &=& - n V T^4 \;{\alpha^2 \over 24} \; {g^4 N^2 \,\lambda_t^2
   \over 48 \pi^4 }\; \Im \quad \mbox{with} \quad \Im =
   \int_0^\infty \! dx\, {x \over e^x -1} \;\int_0^\infty\! dt\,
   {1\over e^{{1\over 2} xt} -1 }\;\cdot\nonumber \\ \label{5end}
 & \cdot & \hspace{-.3cm} 
   \left( {t\over t^2-1} + {4t\over (t^2-1)^2 } + {1\over
   (t+1)^3 } \ln \left( t + 2 \right) + {1\over (t-1)^3 }
   \ln \vert \, t - 2 \vert \right) \;\; . \;\;
\end{eqnarray}
The derivation (a mess) used (\ref{clead1}). To check the
above statement of vanishing soft part, one may write $2/(xt)$ 
for the second Bose function. Then the integral over $t$ 
gives zero, as required. But as it stands, $\Im$ is some 
non--zero mathematical constant ($\Im\approx -1.04$).

The above remaining $\alpha$--dependence, which prevents us 
from solving the gutter problem in the non--Abelian case, is the
''minor detail'' noted in pt.~8 of the Introduction. There must
be a resolution to this puzzle within the truncated version
(\ref{5trunc}), because the terms beyond, depending on $\lambda$,
are of order $g^5$. As the vicious term (\ref{5end}) contains
two Bose functions, the way out has probably nothing to do with
renormalizations. The only possibility we are able to invent is
the fact that at higher orders there is also a ${\sf C}$--term
(see (\ref{2ad})) in the propagator, which is missing in
(\ref{2g}) and is specific to non--Abelian theory. Furthermore,
this term has a factor $\alpha$ in front of it, see e.g. \S ~3
of \cite{flesh}. Let such speculations be beyond the scope of
the present paper.

%
%

\vspace{1.5cm} \centerline{\bf
VI. \ STATIC PROPERTIES } \vspace{1cm} 

So far, while testing the ''even version'' in the 
$\lambda$--case, we were thinking in terms of real 
excitations in the plasma (scalar and gluon), whose 
spectra are hidden in the polarization functions. 
Here we recall the other well--tractable case within the 
infinity of Abelian gauge invariant mass terms.
Before all, turning to the $m$--case comes with a change 
in philosophy. We now ask for the best constant--mass 
terms (longidutinal and transverse) in the trial Lagrangian.
To leading order (otherwise see e.g. \cite{ay}), static 
propagators have the form $(-q^2-m_{\rm screen}^2)^{-1}$. 
But the trial propagators read $(Q_0^2 - q^2 
- m_{t,\ell}^2)^{-1}$. Nevertheless, it may well happen 
(remember the ''$Q_0=0$--method'' of Sec.~IV B) that
they loose memory to their dynamical element $Q_0^2$ 
automatically. 

For Yang--Mills fields, the analysis runs through the steps of
Sec.~V up to (\ref{5vint}). No gauge--fixing dependence occurs
up to the order $g^3$ to be considered here. The bare part
{\gital v}\,$_0$ is given by (\ref{5v0}), now with the $m$--case
integrals (\ref{3rei}), (\ref{3rej}) to be inserted. The crucial
line where the $m$--case starts to make differences reads
\begin{equation} \label{6cruc}
   v_1 = 2T^4 J_0 \left( J_t - J_0 \right)
     - 2 \sum_Q \Delta_{\ell t} (Q) \,{\mit\Pi}_\ell (Q) \;\; .
\end{equation}
Within the present accuracy, the above sum may be reduced
to its $Q_0=0$--term. But note the difference to the
$\lambda$--case. Once the transverse propagator is supplied
with a non--zero magnetic mass by hand, this variational
parameter survives in the result$\,$:
\begin{equation} \label{6soft}
 \sum_Q \Delta_{\ell t} {\mit\Pi}_\ell = { T^4 g^2 N \over
 12\pi}\, \left( \beta m_\ell - \beta m_t \right) \;\; . \;\;
\end{equation}
The same happens in the $J_t$-sum, see (\ref{3rej}). But the
combination of these details in (\ref{6cruc}) yields $v_1= -
T^3 m_\ell /(24\pi)$. The linear (not the cubic, see below)
dependence on $m_t$ has gone, this time by cancellation --
a wanted detail, as we see next. Including the bare part
{\gital v}\,$_0$ the functional reads
\begin{equation} \label{6koma}
   \hbox{{\gital v}\,}^{\,{\rm to\;} g^3}_{\; m{\rm -case}}
   = n\, V\, T^4\, \left( - {\pi^2 \over 45} + {g^2 N \over 144}
  + {1\over 24\pi} \left[ \, (\beta m_\ell )^3 - g^2 N \beta
     m_\ell \,\right]\, + {1\over 12 \pi} (\beta m_t )^3
     \;\right)\;\; . \;\;
\end{equation}
The longitudinal part clearly becomes minimal at $m_\ell = g
\sqrt{N\,}^{\hbox to0.2pt{\hss$
\vrule height 2pt width 0.6pt depth 0pt $}\;\!}   T /
\sqrt{3\,}^{\hbox to0.2pt{\hss$
\vrule height 2pt width 0.6pt depth 0pt $}\;\!}   $,
which is the well known Debye screening mass at leading order.
There is a transverse part in (\ref{6koma}), hence no gutter
problem. As $m_t$ is restricted to the positive half--axis, the
minimum is reached at $m_t=0$, which is the magnetic mass at
the order studied, indeed.

In spite of the above correct answers on static properties,
there remain delicate questions. Remember that the (squared)
Debye mass $3m^2$ already entered the dynamical calculation at
(\ref{4null}). It appears that, within the order $g^3$, the
variational functional can not really discriminate between
statics and dynamics. In fact, the minimum value of the
functional (\ref{6koma}) agrees with (\ref{5min}), i.e. with
the exact one to order $g^3$. Thus, two equally low minima are
found over the space of mass terms. However they are joined,
namely through a subspace of all functions ${\mit \Pi_\ell}$
that have the value $3m^2$ at zero--frequency, and ${\mit
\Pi}_t$ vanishing there. Nevertheless, in the $\lambda$--case
the appearence of constant masses is a technical byproduct,
while in the present static case it answers the posed question.
Let us add conjectures on the behaviour in higher orders. The
safe ground is on the dynamical side. Supplying the variational
functional with anything good then it might answer with
self--energies comparable good. For static properties, on the
other hand, one needs more, namely some philosophy of why the
trial propagators get rid of its dynamical part $Q_0^2$ by only
forcing the mass to be constant. Remember also that, starting
from the real--excitation spectrum in the $\omega$--$q$--plane,
the static limit ($\omega=0$) is only reached through a range
with imaginary wavevector \cite{frosch} on mass--shell lines.
Perhaps, the variational procedure prepares at least the
first non--vanishing term of each screening mass.

At the supersoft scale, the magnetic mass (see \cite{bran,najp}
for more recent work) most probably comes with some numerical
factor times $g^2 T$ \cite{latt}. Then, as a rough speculation,
the last term in
\begin{equation} \label{sepecu}
   \hbox{{\gital v}\,} \; = \; nVT^4 \left( {\rm const}
   + {1 \over 12\pi} (\beta m_t )^3
   - \,{\rm const} \;  g^4 \,\beta m_t \,\right)
\end{equation}
would be in search. Note that such a term, if any and if no
others, would arise in one step over the present truncation of
the functional. For possible danger with this step see the last
point in the following list of open questions.

For completeness, we add the $m$--case result for scalar
electrodynamics. It simply agrees with (\ref{6koma}) at $N=n=1$,
except for the constant terms and an additional term due to the
scalars, which may be both read off from (\ref{43v}). Let us end
up with the question which way the magnetic sectors of Abelian
\cite{blai} and non--Abelian theories might become differenct in
a variational treatment.

%
%

\vspace{1.5cm} \centerline{\bf
VII \ OPEN QUESTIONS } \vspace{1cm} 

In the preceding Sections, the application of the variational
calculus to pure gauge theory was far from being a
straightforward procedure. Several problems were eluded
and questions not answered, because we could not. Let us
recall these questions and just list them here.
\begin{enumerate}
\item  
The Hamiltonian formulation to both, the Gibbs--Bogoljubov
or Feynman--Jensen varational principle (see text below
(\ref{2z})), was given up in Sec.~II because we were unable to
construct the Hamiltonian $H$ of the trial theory. This
construction is a challenging task. See the text below
(\ref{2ama}).
\item  
Knowing the Hamiltonians of both, trial and studied theory,
one could construct the common physical Hilbert space. By
forming the BRST--charge and projecting out physical states
from the outset, this would be the natural approach to the
Gibbs--Bogoljubov version \cite{mank,almu,york}.
\item  
The functional {\gital v}\, in both versions, Gibbs--Bogoljubov
and Feynman--Jensen, has the total minimum value in common
(namely the exact free energy). However, the trial spaces
are different. Hence, a given trial theory which does not
cover this minimum could lead to quite different
approximations. Since presumedly, this is not true, a proof 
of the full equivalence of the two principles is desirable.
Note that such a proof would circumvent our Hamiltonian
problem of the above point 1. Moreover, the interpretation
of the trial space as one of non--equilibrium statistical
operators would be preserved.
\item  
We have not made an effort to introduce, by Legendre 
transformation, the 1PI--generating functional $\Gamma$,
although there is a variational principle even to 
$\Gamma$ \cite{stst,ibpo}.
\item  
Renormalization \cite{mank}, not yet needed in this paper,
is probably inevitable already when the method should
reproduce the next--to leading order perturbative results,
such as e.g. the lowering ''by glue'' of the longitudinal
plasma frequency (for scalar ED this is the term $-0.37e$ 
in eq. (5.5) of \cite{sed}).
\item  
 {}From subsections II B to II C we turned to the ''even
version'' immediately. But perhaps there is something in
between that we have not found, namely a feasible modified
trial theory not running into the pitfall of Sec.~III.
\item  
Only a very poor subspace of polarization functions was 
considered by simply varying prefactors $\lambda_{t,\ell}$
in front of the true functions ${\mit\Pi}_t$, ${\mit\Pi}_\ell$,
already known perturbatively. A honest
''even version''--variational treatment might instead vary
unknown functions ${\mit\Pi}_{t,\ell} (Q)$. To make sense,
this generalization probably needs $g^4$--terms in the
functional {\gital v}\, .
\item  
For the ''minor detail'' of reminescent $\alpha$--dependence
when solving the gutter problem in the non--Abelian case see the
comments at the end of Sec.~V.
\item  
The most terrifying step in Sections IV,V was the expansion
of the $\ln\,\left[\, {\rm ch} (\, )\,\right]\,$ term in the
variational functional. So, the question is whether this
expansion can be avoided some way.
\item  
With regard to the observed gauge--fixing independence, it could
turn out that a later truncation of the series makes less sense
than reading $\ln\left[ {\rm ch} (x)\right]\approx
{1 \over 2} x^2$ as some good approximation.
\end{enumerate}

%
%

\vspace{1.4cm} \centerline{\bf
VIII. \ CONCLUSIONS } \vspace{.9cm}

A Feynman--Jensen type thermal variational principle
is constructed such that an Abelian free trial theory works
well in both cases, scalar electrodynamics and pure
Yang--Mills theory. To this end their actions are to be
rewritten such that only even powers in the gauge field 
appear. This way, the perturbatively known leading--order
self--energies of photons, scalars and gluons, respectively,
are reproduced (apart from a minor open question to the
non--Abelian case) by variation of their prefactors.
The subspace of constant masses covers the inverse
Debye screening length. 
There is a large asymmetry of the functional with
respect to the (photonic/gluonic) transverse sector,
as it does not (yet) depend on the corresponding
parameter at order $g^3$.

The delicate problem of handling two different covariant
gauge--fixing parameters (one of the original and one of
the trial theory) has a simple resolution$\,$: they become
equal by minimization. Hence, the observed
gauge--independence refers to the remaining gauge--fixing
parameter common to both theories.

The new variational functional contains a term    
$\ln \,\left[\, {\rm ch} \left( AAA \right) \,\right]\, $ 
and hence involves arbitrarily high even powers of the gauge 
fields $A$. In the non--Abelian case (and within covariant gauges)
such powers occur already in the unmodified Feynman--Jensen
principle due to the Faddeev--Popov determinant depending on
$A$. Unfortunately, for evaluation and minimization we had to
expand the ln--ch--function. But a true nonperturbative
scheme should never refer to $g$--powers at all. So, the
present success is still below the potential nonperturbative
possibilities of the variational approach.

\vspace{1cm} \centerline{\bf
ACKNOWLEDGEMENTS } \vspace{.3cm}

We are very indebted to Martin Reuter, who made us aware of the
Feynman formulation and correctly localized the origin of our
initial problems in the FP--determinant. We are also grateful
to specific hints and valuable discussions with Norbert Dragon,
Fritjof Flechsig, Edmond Iancu, Olaf Lechtenfeld, Anton Rebhan
and Andreas Wipf.

%
%
\let\tqn=\theequation \renewcommand{\theequation}{A.\tqn}
\setcounter{equation}{0}          

\vspace{1.4cm} \centerline{\bf
Appendix A } \vspace{.8cm}

Here the functional integral measure of the trial partition
function $Z$ of scalar ED is made explicit. $Z$ is (\ref{4zbu})
without the dots there. The normalization factor ${\cal N}$ is
fixed by requiring that, in the massless limit, the partition
function $Z$ must turn into two times that of blackbody
radiation, one of the photons and one of the scalars. On the
more ambitious task of a true derivation see the comments at
the end of this Appendix.

We start by splitting $Z$ into four factors, $Z = Z_\alpha
\, Z_{det} \, Z_A \,Z_s$ with a piece of ${\cal N}$
contained in each. But notice the redundance of such a
factor in front of an unspecified $\int {\cal D} \ldots$,
hence e.g. $Z_\alpha = 1/Z_B\,$ suffices.
The simplest part is $Z_s =$ ''$\!\int {\cal D}
\left\{ \phi^* \phi \right\}$''$\, e^{-S_s}$ with
\begin{equation} \label{asphi}
   S_s = \int^\beta \left( m_s^2 \phi^* \phi + \phi^*
   \partial^2 \phi \right) = \sum_P \left( m_s^2 - P^2 \right)
   \phi(P)^*\phi(P)  = \sum_{{\bf p} \, ,\, n} {m_s^2
   - P^2\over \beta V} \phi^*\phi \;\; . \;\;
\end{equation}
At each of the countable infinite discrete positions ${\bf p}\,
,\, n$ there are, as $\phi$ is complex, two independent
integrations. (\ref{asphi}) refers to our convention
$\phi(x) = \sum e^{-iPx} \phi(P)$ but we may turn to that of
Kapusta \cite{kapu} by $\phi(P) =
\sqrt{{\beta V}\,}^{\hbox to0.2pt{\hss$
\vrule height 2pt width 0.6pt depth 0pt $}\;\!}
  (a + ib) /
\sqrt{2\,}^{ \hbox to0.2pt{\hss$
\vrule height 2pt width 0.6pt depth 0pt $}\;\!}
  $ (with indices ${\bf p}$, $n$ on $a$, $b$ suppressed). We
now guess the functional integral measure and justify by
evaluation$\,$:
\begin{eqnarray} \label{azphi}
  Z_s &=& {\cal N}_0^2 \prod_{{\bf p}\, ,\, n}
  {1\over 2\pi\beta^2} \int \! da\, db \; e^{- {1\over 2}
  \left( m_s^2-P^2 \right) \left(a^2 + b^2\right)} =
  {\cal N}_0^2 \prod \left( - T^2 \Delta_s \right) \;\; ,\;\;
\end{eqnarray}
where $\prod \equiv \prod_{{\bf p}\, ,\, n}$ and
\begin{equation} \label{an0}
  {\cal N}_0 = \prod_{\bf p} \; {\prod_n}^\prime \; 2\pi n \;\;
\end{equation}
with the prime excluding $n=0$. Remember that $\Delta_s^{-1}
= P^2 - m_s^2 = - (2\pi nT)^2 - {\bf p} ^2 - m_s^2 < 0$.
Of course, each factor in ${\cal N}_0$ has to be attached to
the corresponding one in the $n$--product in (\ref{azphi}), and
the product over $n$ has to be performed first (other
constructions may be possible).

The infinite product (\ref{azphi}) can be performed. To this
end we collect four (known) formulas of general use. By contour
integration$\,$:
\begin{equation} \label{af1}
   T \sum_n {1\over P_0^2 - x^2} = - {1\over x} \,\left[\,
   {1\over 2} + n(x) \,\right]\,  \;\;
\end{equation}
with $n(x)\equiv 1 / \left( e^{\beta x} -1\right)$ the Bose
function. (\ref{af1}) is eq. (2.38) of \cite{kapu}. Multiply
(\ref{af1}) with $2x$ and integrate over $x$ from $c$ to $y\,$:
\begin{equation} \label{af2}
   T \sum_n \ln \left( {y^2 - P_0^2 \over c^2 - P_0^2} \right)
   = y - c + 2T \ln \left( {1 - e^{-\beta y} \over 1
   - e^{-\beta c} } \right) \;\; . \;\;
\end{equation}
Multiply (\ref{af2}) with $\beta$, set $\beta y = \omega$ and
perform the limit $\beta c \to 0\,$:
\begin{equation} \label{af3}
  \sum_{n=1}^\infty \ln \left( 1 + { \omega^2 \over
  (2\pi n)^2 } \right)  = \ln \left( {{\rm sh} (\omega /2)
  \over \omega /2} \right) \;\; . \;\;
\end{equation}
Exponentiating (\ref{af3}) and extending to all $n$, one
arrives at the fourth formula
\begin{equation} \label{af4}
  {\prod_n}^{\,\prime}\, {(2\pi n)^2 \over \omega^2 +
  (2\pi n)^2 } = {\omega^2 e^{-\omega} \over \left( 1
  - e^{-\omega} \right)^2 } \;\; , \;\;
\end{equation}
which is eq. (89.5.16) in \cite{hans} and (2.269) in
\cite{klei}. Check (\ref{af4}) at $\omega\to 0$. Using
(\ref{af4}) for (\ref{azphi}) we obtain
\begin{equation} \label{az3}
   Z_s =  {\cal N}_0^2 \prod \left( - T^2 \Delta_s \right)
       = \prod_{\bf p } { e^{-\beta
\sqrt{{m_s^2 + p^2}\,}^{\hbox to0.2pt{\hss$
\vrule height 2pt width 0.6pt depth 0pt $}\;\!}
   } \over \left( 1 - e^{-\beta
\sqrt{{m_s^2 + p^2}\,}^{\hbox to0.2pt{\hss$
\vrule height 2pt width 0.6pt depth 0pt $}\;\!}
   } \right)^2 }  \qquad \mbox{i.e.} \qquad
\end{equation}
\begin{equation} \label{afp}
   F_s = - T \ln \left( Z_s \right) =
   2 \sum_{\bf p} \,\left[\, {1\over 2}
\sqrt{{m_s^2 + p^2}\,}^{\hbox to0.2pt{\hss$
\vrule height 2pt width 0.6pt depth 0pt $}\;\!}
  + T \ln \left( 1 - e^{-\beta
\sqrt{{m_s^2 + p^2}\,}^{\hbox to0.2pt{\hss$
\vrule height 2pt width 0.6pt depth 0pt $}\;\!}
  } \right) \,\right]\, \;\;
\end{equation}
which is, at zero mass, the desired result of twice a half
blackbody radiation. The guessing was good. Aside, one could
include the zero--point energies by the redefinition
${\cal N}_0 \to \prod_{\bf p} e^{\beta p /2}\,
{\prod_n}^\prime\, 2\pi n\,$.

We turn to the factor $Z_{\rm det}$ with again a
trial--and--error prefactor$\,$:
\begin{eqnarray} \label{adet1}
  Z_{\rm det} &=& {\cal N}_0^{-2} {\rm det}^\prime
  \left(\beta^2\partial^2\right) = \,\left[\,  {\cal N}_0^2
  \; {\prod}^\prime \left( - T^2 \Delta_0 \right) \right]^{-1}
    \\ \label{adet2}
 &=& \,\left[\, \; {\prod_{\bf p}}^\prime (\beta p)^2 \; \,\right]\,
  \,\,\left[\, \prod_{\bf p} {\prod_n}^\prime \, { (\beta p)^2
  + (2\pi n)^2 \over (2\pi n)^2 } \,\right]\, \;\; , \;\;
\end{eqnarray}
where in the blank $\prod^\prime$ and on the determinant the
prime excludes only the one position $n = {\bf p} =0$. As the
determinant is the product of the eigenvalues $-\beta^2 P^2$,
naively, $P=0$ must be excluded to make sense. However, if this
is required to result from a derivation, one might go back to
the unity--insertion in the Faddeev--Popov procedure$\,$:
\begin{equation} \label{afadd}
   1 = \Delta \; \cdot \; {\cal N}_0^2 \,\left[\,
   {\prod}^\prime \, T^2 \int da_{{\bf p} ,n} \,\right]\,
   \,\left[\, {\prod}^\prime \delta \left( - P^2 a_{{\bf p} ,n}
   \right)  \,\right]\, \;\; . \;\;
\end{equation}
Originally the $\delta$--argument was $\partial^2 \Lambda$
(with $\delta A_\mu = - \partial_\mu \Lambda$ the gauge
variation). Since there is no constant term in $\Lambda$,
there is no $a_{{\bf 0} , 0}$--integration in (\ref{afadd}) and
no zero $P$ in (\ref{adet1}), q.e.d. ~(\ref{afadd}) directly
leads to $\Delta =Z_{\rm det}$. Using (\ref{af4}) for
(\ref{adet2}) we have
\begin{equation} \label{afdet}
   F_{\rm det} = -T\ln \left( Z_{\rm det} \right) =
   - 2 \; {\sum_{\bf p}}^\prime \,\left[\, {1\over 2} p + T
   \ln \left( 1 - e^{-\beta p} \right) \,\right]\, \;\; . \;\;
\end{equation}
Clearly, with the above measure, the determinant--term subtracts
twice a half blackbody radiation. In passing, the prime in
(\ref{afdet}), while still being necessary there, becomes
irrelevant in the continuum limit.

With an argument quite similar to that below (\ref{afadd}),
there is also a prime in the measure of $\int {\cal D} B$. This
integration runs over a $\delta \left( \partial A - B \right)$.
But $\partial A$ cannot be constant, since otherwise $A$ would
be linear in spacetime and lie outside our space of
Fourier--transformable fields. So, $P=0$ {\sl may} be
excluded$\,$:
\begin{equation} \label{azb}
   Z_B = \hbox{''}\!\!\int {\cal D} B \;\hbox{''}\;
      e^{ - {1\over 2\alpha} \sum_P^\prime B(P)^* B(P)}
       = \prod^{\rm right} {1\over 2\pi} \int da\, db \;
       e^{- {1\over 2\alpha} \left(a^2 + b^2 \right) }
       = \; {\prod}^\prime
\sqrt{\alpha\,}^{\hbox to0.2pt{\hss$
\vrule height 2pt width 0.6pt depth 0pt $}\;\!}  \;\; . \;\;
\end{equation}
As $B(x)$ is a real field and $B(-P)^* =B(P)$, $B(P)=
\sqrt{{\beta V}\,}^{\hbox to0.2pt{\hss$
\vrule height 2pt width 0.6pt depth 0pt $}\;\!}
   (a+ib)/
\sqrt{2\,}^{\hbox to0.2pt{\hss$
\vrule height 2pt width 0.6pt depth 0pt $}\;\!}
   $, the two integrations are placed on half of the
$P$--space, the right say (let right and left exclude the
origin). The prefactor was chosen here to reach the simple
result $ Z_\alpha = 1/Z_B = {\prod}^\prime
\sqrt{{1 / \alpha}\,}^{\hbox to0.2pt{\hss$
\vrule height 2pt width 0.6pt depth 0pt $}\;\!}
   $. It must wait to make sense in combination with $Z_A$.

The photonic part of the trial action includes the mass
terms $M_{t,\ell}\,$:
\begin{eqnarray}
   S_A &=& {1\over 2}\sum_P\left( P^2 - M_t\right) A_\mu^-
   \hbox{\logo A}^{\mu\nu} A_\nu + {1\over 2}\sum_P\left(
   P^2 - M_\ell\right) A_\mu^- {\sf B}^{\mu\nu} A_\nu
      \nonumber \\ \label{aphot}
  & &{} + {1\over 2\alpha}\sum_P P^2  A_\mu^- {\sf D}^{\mu\nu}
  A_\nu \;\;\equiv\;\; S_A^t + S_A^\ell \;\;
\end{eqnarray}
with $A_\mu^- = A_\mu(-P)$ and the ${\sf D}$--term being part of
$S_A^\ell$. The corresponding further splitting $Z_A = Z_A^t
Z_A^\ell$ is allowed because the transverse components (those
in ${\bf e}_{1,2}$--direction, ${\bf e}_{1,2} \perp {\bf p}$,
${\bf e}_1 \perp {\bf e} _2$) in the expansion
\begin{equation} \label{aex}
  A^\mu(P) =  u_1 E_1^\mu + u_2 E_2^\mu + v T^\mu +
              i w U^\mu  \quad \mbox{with} \quad
  T^\mu = \left( 0, {{\bf p} \over p}\right) \;, \;
  E_{1,2}^\mu = \left( 0,{\bf e}_{1,2} \right) \;\; , \quad
\end{equation}
drop out in $S_A^\ell$ and are the only parts surviving under
the $\hbox{\logo A}$--operation$\,$: $\hbox{\logo A}^{\mu\nu}
E_{1,2\;\nu} = E_{1,2}^\mu$. As the first three terms of
(\ref{aex}) as well as $wU^\mu$ are Fourier transforms of real
fields, half $P$--spaces are related by $u_j(-P) E_j^\mu(-P)
= u_j(P)^* E_j^\mu(P)$, $v(-P) = - v(P)^*$ and $w(-P)=w(P)^*$.
Hence, the integrations in $Z_A^t$, to start with, are of the
real--field type (\ref{azb}), except that there are now two
integrations at the origin $n={\bf p} =0$ and four at each place
in the right half. Two of the latter may be attached with the
left half. Then, choosing the same functional integral measure
as for $Z_s$, we arrive at precisely (\ref{azphi}) with the
role of $m_s^2$ taken over by $M_t(P)\,$:
\begin{equation} \label{azat}
   Z_A^t = {\cal N}_0^2 \;\prod \left( - T^2 \Delta_t
   \right) \;\; . \;\;
\end{equation}
The longitudinal part of the action is first rewritten as
\begin{equation} \label{aphol}
   S_A^\ell = {1\over 2} {1\over \beta V} \sum_{{\bf p} ,
   n}^{\rm all} \left( \,\left[\, M_\ell - P^2 \,\right]\,
   \vert \xi \vert^2 - P^2 {1\over \alpha} \vert \eta \vert^2
   \right) \quad \mbox{with}  \quad \matrix{ \xi = p w
   + i P_0 v \cr \eta = -iP_0 w + p v \cr } \;\; . \;\;
\end{equation}
Next we observe that $\xi (-P) = \xi (P)^*$, $\eta (-P) = - \eta
(P)^*$, and mark the origin and the right half $P$--space to
count independent integrations (two over $\xi$ at the origin and
four in the right). Finally, by changing the variables from $v$,
$w$ to $\xi$, $\eta$ (with unit Jacobian determinants), and with
the now familiar functional integral measure, one arrives at
\begin{equation} \label{azal}
   Z_A^\ell = {\cal N}_0 \;\prod
\sqrt{{ - T^2 \Delta_\ell}\,}^{\hbox to0.2pt{\hss$
\vrule height 2pt width 0.6pt depth 0pt $}\;\!} \;
   {\cal N}_0 \;{\prod}^\prime
\sqrt{{ - T^2 \alpha \Delta_0 }\,}^{\hbox to0.2pt{\hss$
\vrule height 2pt width 0.6pt depth 0pt $}\;\!} \;\; . \;\;
\end{equation}
Note that most of the above ''trivialities'' were due to
carefully counting all positions in $P$--space, i.e. to
place the primes right. 

We are ready to constitute the scalar ED partition function 
from the above several factors$\,$:
\begin{eqnarray}    \hspace{-.3cm}
   Z &=& {1\over \; {\prod}^\prime
\sqrt{\alpha\,}^{ \hbox to0.2pt{\hss$
\vrule height 2pt width 0.6pt depth 0pt $}\;\!}
   } \;\; {1\over {\cal N}_0^2 \; {\prod}^\prime \left(
   - T^2 \Delta_0 \right) } \;\; {\cal N}_0^2 \prod \left(
   - T^2 \Delta_t \right) \; \cdot \nonumber \\ \label{aend}
   &\cdot& \!\!
\sqrt{{{\cal N}_0^2 \prod \left( -T^2 \Delta_\ell \right)}
   \,}^{\hbox to0.2pt{\hss$
\vrule height 2pt width 0.6pt depth 0pt $}\;\!}   \;\;
\sqrt{{ \,\left[\, {\prod}^\prime \alpha \,\right]\,  \;
   {\cal N}_0^2 \; {\prod}^\prime \left( -T^2 \Delta_0 \right) }
   \,}^{\hbox to0.2pt{\hss$
\vrule height 2pt width 0.6pt depth 0pt $}\;\!} \;\;
   {\cal N}_0^2 \prod \left( -T^2 \Delta_s \right) \;\; . \quad
\end{eqnarray}
Obviously, the gauge--fixing parameter $\alpha$ cancels. Now,
counting halves of blackbody radiation amounts to 
$ - 2 + 2 + 1 + 1 + 2 = 4$ as required.

A true derivation of the above must not anticipate the known
zero--mass results. With \cite{bern} as a guideline, such
derivation should be possible even inside covariant gauges,
i.e. without a recourse to physical gauges. There is one
problem in taking the right starting point (maybe with a
factor ${\cal N}_0^4$ in front of the classical partition
function for the four [of six] degrees of freedom to be
quantized), and in the volume factor (to be split off) the
other.

%
%
\renewcommand{\theequation}{B.\tqn} \setcounter{equation}{0}
   
\vspace{1.5cm} \centerline{\bf   
Appendix B } \vspace{.8cm}

Here the two sums $L_t$ and $L_t$ are evaluated, in the
$\lambda$--case and with regard to contributions not accessible
by a naive $Q_0=0$--method. The details are required for
subsections IV B and IV D. We start from the definition
(\ref{3sums}) and work with the spectral representation
\begin{equation} \label{4rhodef}
  {1 \over P^2 - \lambda^2 {\mit \Pi}_{t,\ell} (P)}
  = \int\! dx \, x \, {\rho^{(\lambda )}_{t,\ell}
  (x , p ) \over P_0^2 - x^2 }\;\; , \;\;
\end{equation}
of trial propagators. The above spectral densities are related
to ordinary ones, denoted by $\rho_{t,\ell} (x,p\, ; m^2)$, by
\begin{equation} \label{4rhol}
  \rho_{t,\ell}^{(\lambda)} (x,p) = \rho_{t,\ell}
  \left( x,p\, ; \lambda^2 m^2 \right) \;\; . \;\;
\end{equation}
Hence all sum rules (\ref{csumt}), (\ref{csuml}) remain
valid for $\rho^{(\lambda)}$ if $m^2$ is replaced by
$\lambda^2m^2$ to the right. Using (\ref{4rhodef}) and the sum
rule $1 = \int\! dx \, x\, \rho^{(\lambda )} (x,p)$ the
$L$--sums read
\begin{equation} \label{b1}
  L_{t,\ell} = - \beta^4 \sum_P \left( P^2 \Delta_{t,\ell}
  - 1 \right) = - \beta^4 \sum_P \int\! dx\, x \,
  \rho_{t,\ell}^{(\lambda )} (x,p) \left( { P_0^2 - p^2
  \over P_0^2 - x^2 } - 1 \right) \;\; . \;\;
\end{equation}
Next, with (\ref{4rhol}), defining $\overline{\rho} \equiv
(x^2-p^2) \,\rho$ (cf. (\ref{cquer})) and using (\ref{af1}),
we may write
\begin{equation} \label{b2}
    L_{t,\ell} = \beta^4 {1\over 2\pi^2 }\int_0^\infty\!
    dp \, p^2 \int\! dx \;\overline{\rho}_{t,\ell}
    (x,p\, ; \lambda^2 m^2 ) \,\,\left[\, {1\over 2}
    + n(x) \,\right]\, \;\; . \;\;
\end{equation}
Note that both, $\overline{\rho}$ and the square bracket, are
odd functions of $x$. Let us split $\overline{\rho}$ into its
leading part as given by (\ref{clead1}), (\ref{clead2}), and
the rest $\overline{\rho} - \overline{\rho}^{\;\rm lead}$,
which we call the soft part. Correspondingly, $L$ is written
as $L^{\rm lead} + L^{\rm soft}$. Introducing an UV--cutoff
$\Lambda$, the leading parts may be written as
\begin{equation} \label{b3}
   \left. \matrix{ L^{\rm lead}_t \cr L^{\rm lead}_\ell \cr}
   \right\} = \left\{ \matrix{ \lambda_t^2 \cr 2
   \lambda_\ell^2 \cr } \right\} {\beta^2 g^2 \over 12\pi^2}
   \int_0^\Lambda\! dx\, x \,\left[\, {1\over 2} + n(x)
   \,\right]\, \left\{ \matrix{ 1 - I(x) \cr I(x) \cr } \right.
   \; , \;\; I(x) = \int_x^\Lambda\! dp\, {1\over p} \;\; .\;\;
\end{equation}
(\ref{b3}) is the right place deleting the ''${1\over 2}$--term''
as discussed below (\ref{3j}) in the main text. But even under
the control of the Bose function there remains a logarithmic
divergent factor, namely
\begin{equation} \label{b4}
  \kappa = {\beta^2 \over 4\pi^2} \int_0^\Lambda \! dx\, x \,
  n(x) \int_0^\Lambda \! dp\, {1\over p} \;\; . \;\;
\end{equation}
Using (\ref{b4}) we have $L^{\rm lead}_t = \left( {1\over 24}
- \kappa \right) {1\over 3} g^2 \lambda_t^2 $ and
$L^{\rm lead}_\ell = {2 \over 3} \kappa g^2 \lambda_\ell^2 $,
which are the $\lambda^2$--terms in (\ref{4lex}).

We turn to the soft parts of $L_{t,\ell}$, whose series might
start with $g^2\lambda^3$. To prepare this $\lambda^3$--term,
one may simply write $T/x$ in place of the square bracket in
(\ref{b2}) (and, of course, the mentioned difference in place
of $\overline{\rho}\,$). Using the sum rule (\ref{cssuml}), one
obtains
\begin{equation} \label{b5}
   L_\ell^{\rm soft} = {\beta^3 \over 2\pi^2} \int_0^\infty\!
   dp \, p^2 \int\! dx \, {1\over x} \,\left[\,
   \overline{\rho}_\ell (x,p)  - \overline{\rho}_\ell^{\;\rm
   lead} (x,p) \right]_{m^2 \to \lambda^2 m^2}
   = - {1\over 4\pi} \left( {g \lambda_\ell \over
\sqrt{3\,}^{\hbox to0.2pt{\hss$
\vrule height 2pt width 0.6pt depth 0pt $}\;\!}
   } \right)^3 \;\; . \;\;
\end{equation}
But, through the above line and with view to (\ref{cssumt}), the
transverse function $L_t^{\rm soft}$ vanishes. This completes
the derivation of (\ref{4lex}).

For the gutter problem of Sec.~IV D we must still learn
about the first non--vanishing piece of $L_t^{\rm soft}$. Let
us work with $\lambda_t=1$ and remember $m\to\lambda_t m$ at
the end. We start from the full expression, but separate the
cut and pole parts of the spectral densities. In particular,
$\overline{\rho}_t^{\,\rm cut,\, lead}$ means the second term
in (\ref{clead1}), and $\overline{r}_t^{\;\rm lead} = 3m^2 /4p$
the prefactor of the delta functions. There is an exact
expression (without index lead) to both. Then, three
differences may be formed$\,$:
\begin{eqnarray}
   L_t^{\rm soft} (\lambda_t=1) &=& {\beta^4 \over 2\pi^2 }
   \int_0^\infty\! dp \, p^2 \int\! dx \,\left[\, {1\over 2}
   + n(x) \,\right]\, \left\{ 2 \left( \overline{r}_t
   - \overline{r}_t^{\;\rm lead} \right) \delta (x - \omega_t )
   \right. \nonumber \\ \label{b6}
 & & \left. +\;\,\overline{\rho}_t^{\,\rm cut}
   - \overline{\rho}_t^{\,\rm cut,\, lead}
   + 2 \overline{r}_t^{\;\rm lead}\,\left[\,\delta (x-\omega_t )
   - \delta (x-p) \,\right]\, \right\} \;\; , \;\;
\end{eqnarray}
where $\omega_t = \omega_t (p)$ is the transverse plasma
frequency, to be obtained by solving $\omega_t^2 = p^2
+ {\mit\Pi}_t (\omega_t,p)$. We now notice that $x$, $p$ are
restricted to soft values by the above first two differences,
but not by the third one. So, in front of the first two, we
may still use the $T/x$ approximation. Note that ${1\over 2}
+ n(x) -T/x = \beta x / 12 + O(\beta^2 x^2)$. Hence, for
$\beta x \sim g$ this difference is by two $g$--powers smaller
than $T/x \sim 1/g$. It might contribute to $L$ only at $g^5$.
Working this way, the sum rule helps again to get rid of
$\overline{\rho}_t$ and $\overline{r}_t\;$:
\begin{equation} \label{b7}
   L_t^{\rm soft} (\lambda_t=1) = {\beta^3 \, 3 m^2
   \over 4\pi^2 } \int_0^\infty\! dp \, \left( 1
   - {p\over \omega_t }  + \beta p \,\left[\, n \left(
   \omega_t\right) - n(p) \,\right]\, \right)  \;\; . \;\;
\end{equation}
For convenience, this can be further rewritten by introducing 
$\omega = \omega_t$ as the integration variable (and by once
more replacing $n(p) \to T/p$ in a soft term -- this time
required for consistency)$\,$:
\begin{equation} \label{b8}
   L_t^{\rm soft} (\lambda_t=1)  = { \beta^3 \, 3m^2
   \over 4\pi^2 } \int_m^\infty \! d\omega \,\left[\,  1
   - \beta \omega n(\omega ) \,\right]\, \left( 1 - {p(\omega )
   \over \omega \omega^\prime} \right) \;\;
\end{equation}
with $\omega^\prime$ the derivative of $\omega_t$ with respect
to $p$, and $p$ being $p(\omega )$. The square bracket starts
as ${1\over 2} \beta\omega$ for small $\beta\omega$, its
saturation at 1 being never reached because the round bracket
sets the limit. It starts with 1/6 (at $\omega \to m$) and goes
as $(9/4) m^4 \omega^{-4} \ln(\omega /m)$ for large $\omega$
(with such details taken from Appendix B of \cite{nt}). Hence
(\ref{b8}) is indeed of order $g^4$ in magnitude. Going to
$\lambda_t\neq 1$ simply amounts to $m \to \lambda_t m $ in
(\ref{b8}). But note that this scaling also changes the
definition of e.g. $\omega_t\,$, which now is the
transverse plasma frequency as if $m$ were $\lambda_t m$.

What we really need in the main text, is not $L_t$ itself 
but the first factor in (\ref{44dw}). The operation there,
fortunately, eliminates the above last integration$\,$:
\begin{equation} \label{b9}
  {2\over \lambda_t} L_t - \partial_{\lambda_t}\, L_t \; = \;
  {g^3 \lambda_t^2 \over 216 \pi^2 } \; \,\left[\, 1
  - {g\lambda_t\over 3} \, n \left( {T g \lambda_t \over 3 }
  \right) \,\right]\, \;\; . \;\;
\end{equation}
This ''first factor'' is thus positive, and it behaves as
$\sim g^4 \lambda_t^3$ for small $g$. Just these properties
were used in the main text below (\ref{44dw}) to reach the
long bath--tub.

%
%
\renewcommand{\theequation}{C.\tqn} \setcounter{equation}{0}
   
\vspace{1.5cm} \centerline{\bf   
Appendix C } \vspace{.8cm}

Here we collect a few special details on the spectral densities
$\rho_t$ and $\rho_\ell$ which were needed in Appendix B. There
we had to learn on the product
\begin{equation} \label{cquer}
   \overline{\rho}\; (x,p) \;\;\equiv\;\; ( x^2 - p^2 )
   \;\rho\, (x,p) \;\; 
\end{equation}
and its asymptotic forms at large $p$--argument
($p^2 \gg m^2 $)$\,$:
\begin{eqnarray} \label{clead1}
   \overline{\rho}_t^{\;\;\rm lead} \;\; &=& \;\; {3m^2
   \over 4p} \,\left[\, \delta (x-p) - \delta (x+p) \,\right]\,\;
    - \; {3m^2\over 4 p^3}\, x \; \theta \left( p^2-x^2 \right)
    \\  \label{clead2}
   \overline{\rho}_\ell^{\;\;\rm lead} \;\; &=& \;\; + \; {3m^2
   \over 2 p^3}\, x \;\theta \left( p^2-x^2 \right) \quad . \;\;
\end{eqnarray}
These leading terms are readily obtained from the full
expressions as given in Appendix B of \cite{nt}. One may check
(\ref{clead1}) and (\ref{clead2}) by using it in the
$\overline{\rho}$ sum rules and thereby producing, in each case,
the term of highest $p\,$--power to the right. The exact
$\overline{\rho}$ sum rules read$\,$:
\begin{equation} \label{cssumt}
   \int \! dx \, \left\{
 \matrix{ \raise 11pt\hbox{}
   1/x \lower 8pt\hbox{} \cr  x \lower 8pt\hbox{} \cr
   x^3 \lower 8pt\hbox{} \cr  x^5 \lower 8pt\hbox{} \cr
   x^7 \lower 11pt\hbox{} \cr } \right\}
     \overline{\rho}_t (x,p) \;\; = \;\; \left\{
 \matrix{ \raise 11pt\hbox{} 0 \lower 8pt\hbox{} \cr
   m^2 \lower 8pt\hbox{} \cr
   {6\over 5} p^2 m^2 + m^4 \lower 8pt\hbox{} \cr
   {9\over 7} p^4 m^2 + {12\over 5} p^2 m^4 + m^6
       \lower 8pt\hbox{} \cr
   {4\over 3} p^6 m^2 + {702\over 175} p^4 m^4 + {18\over 5}
    p^2 m^6 + m^8 \quad , \lower 11pt\hbox{} \cr } \right. \;\;
\end{equation}
\begin{equation} \label{cssuml}
   \int \! dx \, \left\{
 \matrix{ \raise 11pt\hbox{}
   1/x \lower 8pt\hbox{} \cr   x \lower 8pt\hbox{} \cr
   x^3 \lower 8pt\hbox{} \cr   x^5 \lower 8pt\hbox{} \cr
   x^7 \lower 11pt\hbox{} \cr} \right\}
     \overline{\rho}_\ell (x,p) \;\; = \;\; \left\{
 \matrix{ \raise 11pt\hbox{} 3m^2 / \left( 3m^2 + p^2 \right)
    \lower 8pt\hbox{} \cr   m^2 \lower 8pt\hbox{} \cr
    {3\over 5} p^2 m^2 + m^4 \lower 8pt\hbox{} \cr
    {3\over 7} p^4 m^2 + {6\over 5} p^2 m^4 + m^6
       \lower 8pt\hbox{} \cr
    {1\over 3} p^6 m^2 + {213\over 175} p^4 m^4 + {9 \over 5}
    p^2 m^6 + m^8 \quad . \lower 11pt\hbox{} \cr } \right. \;\;
\end{equation}
They derive through (\ref{cquer}) from the sum rules of ordinary
densities$\,$:
\begin{equation} \label{csumt}
   \int \! dx \, \left\{
 \matrix{ \raise 11pt\hbox{} 1/x \lower 8pt\hbox{} \cr
   x \lower 8pt\hbox{} \cr    x^3 \lower 8pt\hbox{} \cr
   x^5 \lower 8pt\hbox{} \cr  x^7 \lower 8pt\hbox{} \cr
   x^9 \lower 11pt\hbox{} \cr } \right\}
    \rho_t (x,p) \;\; = \;\; \left\{
 \matrix{ \raise 11pt\hbox{} 1/p^2 \lower 8pt\hbox{} \cr
   1 \lower 8pt\hbox{} \cr     p^2 + m^2 \lower 8pt\hbox{} \cr
   p^4 + {11\over 5} p^2 m^2 + m^4 \lower 8pt\hbox{} \cr
   p^6 + {122\over 35} p^4 m^2 + {17\over 5} p^2 m^4
      + m^6 \lower 8pt\hbox{} \cr
   p^8 + {506\over 105} p^6 m^2 + {1297\over 175} p^4 m^4
      + {23\over 5} p^2 m^6 + m^8
      \quad , \lower 11pt\hbox{} \cr }  \right. \;\;
\end{equation}
\begin{equation} \label{csuml}
   \int \! dx \, \left\{
 \matrix{ \raise 11pt\hbox{} 1/x \lower 8pt\hbox{} \cr
   x \lower 8pt\hbox{} \cr     x^3 \lower 8pt\hbox{} \cr
   x^5 \lower 8pt\hbox{} \cr   x^7 \lower 8pt\hbox{} \cr
   x^9 \lower 11pt\hbox{} \cr } \right\}
     \rho_\ell (x,p) \;\; = \;\; \left\{
 \matrix{ \raise 11pt\hbox{} 1 / \left( 3m^2 + p^2 \right)
     \lower 8pt\hbox{} \cr  1 \lower 8pt\hbox{} \cr
   p^2 + m^2 \lower 8pt\hbox{} \cr
   p^4 + {8\over 5} p^2 m^2 + m^4 \lower 8pt\hbox{} \cr
   p^6 + {71\over 35} p^4 m^2 + {11\over 5} p^2 m^4
       + m^6 \lower 8pt\hbox{} \cr
   p^8 + {248\over 105} p^6 m^2 + {598\over 175} p^4 m^4
               + {14\over 5} p^2 m^6 + m^8
     \quad , \lower 11pt\hbox{} \cr }   \right. \;\;
\end{equation}
and these, in turn, are derived along the lines
given in \cite{sumr}.

%
%

    \vspace{.5cm}   \renewcommand{\section}{\paragraph}
    
\end{document}